\documentclass[11pt,a4paper]{article}
\usepackage[T2A]{fontenc}
\usepackage{amsmath,amssymb}
\usepackage{epsfig} %

\begin{document}

\bigskip

\begin{center}
\textit{\bf DENSITY-WAVE SPIRAL THEORIES IN THE 1960s. I.}
\end{center}

\bigskip

\begin{center}
\textbf{I. I. Pasha}\footnote{ \textit{ ${} $ email}:
misterwye@mtu-net.ru} \par
\bigskip
{\footnotesize \textit{ Sternberg State Astronomical Institute,
Moscow} \par \textit{ Vavilov Institute for History of Science
and Technology, Moscow }}

\bigskip
\bigskip
\bigskip

\tableofcontents

\end{center}

\newpage

{\footnotesize \begin{list}{}{\leftmargin4cm}
\item You men are strange people -- said Amaranta, unable to think up anything
else. -- All your life you fight against priests, but give prayer-books.
\begin{flushright}
\textit{G.G. Marquez. A hundred years of solitude}
\end{flushright}

\item \ldots  a common foible of those who in the feeling of devotion are disposed
to exaggerate the significance of their heroes.
\begin{flushright}
\textit{Einstein 1953}
\end{flushright}
\end{list}}

\bigskip

\section*{Introduction\protect\footnotemark{}}
\addcontentsline{toc}{section}{Introduction}
\footnotetext{ Throughout the paper, the \textit{italicized} names in parentheses
refer to private communications as identified in the note to the
list of references.}

\bigskip

\noindent The modern density-wave theory of spiral structure in
galaxies, sprung in the 1960s, had long been preceded by the
theories of Bertil Lindblad. Those started back in the days when
Hubble demonstrated that whirlpool nebulae reside far outside the
Galaxy, and when Jeans conveyed an engrossing feeling of steady
spirals ordered by yet unknown forces.\footnote{ ``Each failure to
explain the spiral arms makes it more and more difficult to
resist a suspicion that the spiral nebulae are the seat of types
of forces entirely unknown to us, forces which may possibly
express novel and unsuspected metric properties of space. The
type of conjecture which presents itself, somewhat insistently,
is that the centres of the nebulae are of the nature of `singular
points', at which matter is poured into our universe from some
other, and entirely extraneous, spatial dimension, so that, to a
denizen of our universe, they appear as points at which matter is
being continually created'' (Jeans 1929, p.360).} Astronomer by
education, Lindblad did not yield to temptation by this imposing
obscurity of force, and he headed for a dynamical spiral theory
in terms of ordinary gravitation.\footnote{ Polemizing with Jeans
on the spiral problem, Brown, a celestial mechanician from Yale
University, defended already its gravitational status. In his
mind, star orbits might at certain conditions correlate in shape
and orientation so as to reveal a two-armed spiral-like envelope,
thus delineating a ``visible structure [\ldots ] due to the
greater space density of visible matter in the neighborhood of
the arms than elsewhere'', i.e. a \textit{stationary wave of
condensation} (Brown 1925, p.109-10). Noticed though (Jeans 1929;
Lindblad 1927c), Brown's work had no perceptible impact. \par
\par} Right then, this task must have appeared extremely
difficult, to be at best a matter of a lifetime of work, since
the analytical methods of the patronizing disciplines
(hydrodynamics, statistical mechanics) were rudimentary and gave
almost nothing for the stellar-dynamical research. Still more
striking was Lindblad's break-through in the field of stellar
kinematics. By 1927 already he developed the theory of epicycles,
having shown that a star moving on a nearly circular galactic
orbit just oscillates about its mean radius (Lindblad 1926b). The
frequency $\kappa $ of such oscillations was given by the
relations

\bigskip

\begin{equation}
\kappa / 2\Omega =
\left( {1 - {{A} \mathord{\left/ {\vphantom {{A} {\Omega}} } \right.
\kern-\nulldelimiterspace} {\Omega}} } \right)^{1 / 2} = c_{\theta}  / c_{r}
\end{equation}

\bigskip

\noindent
including the angular speed $\Omega $, the Oort constant of differential
rotation $A \equiv - \raise.5ex\hbox{$\scriptstyle 1$}\kern-.1em/
\kern-.15em\lower.25ex\hbox{$\scriptstyle 2$} rd\Omega / dr$, and the
azimuthal-to-radial velocity dispersion ratio (Lindblad 1927b); the values
of $c_{\theta}  / c_{r} $ got remarkably close as calculated and empirically
determined for the solar neighborhood (Lindblad 1929). These results
reinforced the stellar-dynamical foundations and also they gave Lindblad
confidence in his search of the origins and mechanisms of the galactic
spiral phenomenon, but, quickly recognized and instigated by success, he was
taken hostage, then and on, to the epicyclic-orbit scheme.

\bigskip
\bigskip

\section*{I. LINDBLAD'S ERA}
\addcontentsline{toc}{section}{I. Lindblad's era}

\bigskip

{\footnotesize \begin{list}{}{\leftmargin4cm}
\item The only result that seems to emerge with some clearness is that the spiral
arms are permanent features of the nebulae [...] perpetuated in
static form.
\begin{flushright}
\textit{Jeans 1929, p.360}
\end{flushright}
\end{list}}

\bigskip

\subsection*{1.1 From unstable orbits to global wave modes}
\addcontentsline{toc}{subsection}{1.1 \it From unstable orbits to global wave modes}

\bigskip

{\footnotesize \begin{list}{}{\leftmargin4cm}
\item It is natural that in this field, on which at that time nothing was ripe for
harvesting, he did not immediately find the right path.
\begin{flushright}
\textit{Oort 1967, p.333}
\end{flushright}
\end{list}}

\bigskip

\noindent Though the fact of our larger-scale universe had begun to emerge through
Hubble's work, it was not yet as clear on the quantitative side: well
advanced in rank, the `nebulae' still came short of size and mass against
our Galaxy. This was made by the underrated galaxy-distance scale,\footnote{
It was not until the early 1950s that the distance scale was reconsidered
(see Baade 1963, Efremov 1989) and the size of the Local Group doubled.
Given the shifted zero-point in the Cepheid-luminosity calibration, Hubble's
constant was reduced, and by the 1960s it fell from its original 550
\textit{km/s/Mpc} down to 180 (de Vaucouleurs) or to 80 (Sandage). This gave a 3-to-7-fold
increase in distance.} and the giant ellipticals, missing in the Local Group
and nearby, got it the most. On the whole, the ellipticals were found to be
one to two orders under the spirals, and the rather enigmatic barred
galaxies were ranged somewhere intermediate (Hubble 1936).

\bigskip

Original absorption-spectrum methods of detecting the galaxy rotation were
sensitive only for bright central regions of comparatively close systems,
the line inclination being established integrally, as a quantitative measure
of overall uniform rotation. The emission-spectrum methods, in practice since
the late 1930s, could as well catch the kinematics of the rather distant
regions in our next-door spirals M31 and M33 (Babcock 1939, Mayall {\&}
Aller 1942). Limited and inaccurate though these data were (Fig.1), they
took astronomers by storm and for almost two decades then they formed and
served the idea of a standard rotation curve. The latter was understandably
professed to obey $V(r) = a{{r} \mathord{\left/ {\vphantom {{r} {(1 +
br^{2})}}} \right. \kern-\nulldelimiterspace} {(1 + br^{2})}}$ and be scaled
so as to co-measure its rising part to a live galaxy within its `visible
boundary'.\footnote{ This form of $V(r)$ emerged from the solution of Jeans'
problem for an axisymmetric stationary stellar system with ellipsoidal
velocity distribution. It greatly encouraged work on modeling the
three-dimensional gravitational potential and mass distribution in the
Galaxy (Parenago 1950, 1952; Kuzmin 1952; Safronov 1952; Idlis 1957).}$^{,
}$\footnote{ ``Both in M31 and M33 the easily visible spiral arms lie in
regions where the rotation does not deviate strongly from uniformity. It is
remarkable in M31 that outside the nucleus [\ldots ] there is another region of
nearly uniform rotation'' (Weizsacker 1951, p.179). Vorontsov-Velyaminov
(1972) was still confident that near uniform rotation was the type adopted
by most of spiral galaxies.} $^{} $And on the barred spirals it was
disarmingly clear ``with no measurement'' at all that in face of rapid bar
destruction their rotation was nothing, if not uniform (Ogorodnikov 1958, p.517).

\begin{figure}
\centerline{\epsfxsize=0.9\columnwidth\epsfbox{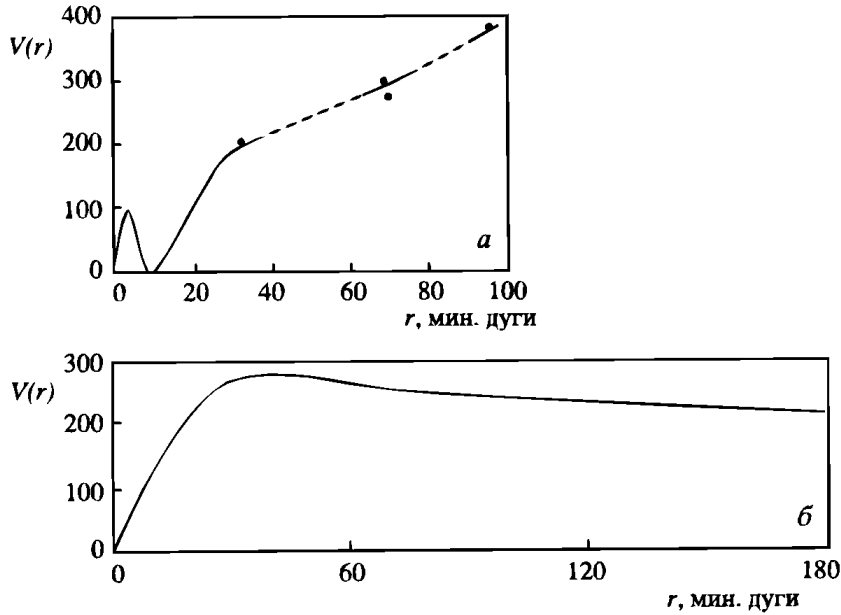}}
\caption{\footnotesize \textit{The rotation curve of the galaxy M31}:
$a$ -- as provided by the late-1930s optical data (Babcock 1939), $b$ --
as inferred from the mid-1950s radio data (Hulst et al 1957).}
\end{figure}

\bigskip

Genuinely matched with the empirical climate were the theoretical tastes of
the epoch that followed closely Jeans' directive on unified cosmogony of
galaxies and stars.\footnote{ The idea of an overall one-time star formation
early in the life of our Galaxy had long been predominant. In the late 1930s
only the hydrogen-to-helium-synthesis energy source was proposed. That
allowed evaluation of the fuel exhaustion time at a given star luminosity,
and its shortness for the blue supergiants -- 10$^{7}$ yrs -- exhibited star
formation as an ongoing process. This idea gained empirical support during
the 1940s.} One relied on the study of gaseous figures; they were diagnosed
to be open to evolutive secular instability created by dissipation factors
acting in the steady-motion systems. The latter just ``never attain to a
configuration in which ordinary [dynamical] instability comes into
operation'' (Jeans 1929, p.199), so that ``it is secular stability alone
which is of interest in cosmogony'' (Jeans 1929, p.214)\footnote{ In Jeans'
view (Jeans 1929, p.213), as a nebula in uniform rotation shrinks, it
alters (augments) density, not angular momentum, running through a
one-parameter sequence of equilibrium figures. Remarkably, this same
sequence is followed by a non-compressible liquid body as it enhances its
momentum. According to Poincare, this body is secularly stable till it is a
low-flattening Maclaurin spheroid. But when some critical eccentricity
(momentum) is reached, it looses stability, takes another sequence of stable
equilibrium figures -- Jacobi ellipsoids -- and then follows it at speedier
rotations.} . Quite understandably, Lindblad's early work lay nearby in the
feeling for global evolutionary processes.\footnote{ ``Now it is obvious
from the scheme as Hubble described it that he had an impression or a
belief, although he never quite admitted it, that it represented a
continuous sequence. But I believe, on the contrary, that Lindblad put his
finger on the essence of Hubble's classification when he suggested that it
is a series of increasing flattening, or increasing angular momentum''
(Baade 1963, p.16-17). \par ``According to Lindblad's theory, the fully
resolved spiral pattern is regarded as an advanced state which all nebulae
will eventually reach in the course of their evolution'' (Chandrasekhar
1942, p.180).} Yet he was the first, and for more than thirty years almost
the only one, who singled out the spiral problem and treated it as a
separate, \textit{stellar-dynamical} element in the general philosophy of galaxies.\footnote{ The
trend of this philosophy is sensed through the following reflection by
Weizsacker (1951, p.165): ``The evolution of a single object can be
understood only if its temporal and spatial boundary conditions and the
external forces acting on it are known. These are defined by the evolution
of the larger system of which the object forms a part. So every single
problem is likely to lead us back into the problem of the history of the
universe''.}

\bigskip

Lindblad started from a highly flattened lens of stars in uniform rotation
($\Omega = const, A = 0$ in Eqn (1)) created in the course of primary
evolution (Lindblad 1926a, 1927a). Gravitational potential at its edge
changes so abruptly with radius that circular orbits there get unstable
($\kappa ^{2}$ < 0): those inside of, but close to, the edge need only a
slight individual change in energy in order to be transformed into
quasi-asymptotic orbits extending very far from the `mother system' (the
solar neighborhood belongs exactly to some such exterior that shows
differential rotation obeying relations (1)). Still stars leave and return
to their mother system spontaneously and equiprobably in any point on its
edge, which is not conducive to neat global patterns. But the hitch is
removed upon the admission of either an outside disturber or an overall oval
distortion caused by fast rotation.\footnote{ Circular orbits at the
spheroidal edge are unstable for eccentricities $ e_{1} $> 0.834, and as the
level $e_{2}$ = 0.953 is achieved (3.1:1 axis ratio), dynamical instability
against the two-crest harmonic sectorial waves is thrown in, so that the
figure gets oval.} In both cases, two opposite ejection points arise on the
edge of the lens after a transitory process and, fixed in space, they pour
material out in spiral-looking leading gushes. Turning to \textit{intrinsic} mechanisms of
galaxy structures, Lindblad laid greatest stress upon global modes of
disturbances, called the deformation waves (`uncompressible' modes) and the
\textit{density waves} (`compressible' modes), and sought their unstable solutions
(Fig.2).\footnote{ ``The most important modes of density variation'' appear to be
of the type of $\sim (r / R)^{m}\cos (\omega t - m\theta )$ ($\omega $
and $m$ being wave frequency and azimuthal wavenumber, $R$ -- the lens radius).
``The conditions for instability have been investigated for the waves $m = $1, 2,
3. The greatest interest attaches to the wave $m = $2 because it tends to explain
the formation of barred spirals. The density variation is accompanied by the
development of four whorl motions. [\ldots ] The disturbances due to the
four whorls on the motions in a surrounding ring structure [the latter
thought of as having been formed previously] explain in a qualitative way
the development of spiral structure'' (Lindblad 1962, p.147).} Analyzing
the effects such waves had on stars on asymptotic orbits (Fig.3), he
proposed and refined scenarios of spiral-arm formation in an outer, shearing
galaxy envisaged to keep up somehow the patterns as arranged by a mass of
the affected orbits, rather than to destroy them (Lindblad 1927a, 1948,
1953).\footnote{ These articles provide a reasonable summary of Lindblad's
theories prior to 1955. The asymptotic-spiral theory was thoroughly reviewed
by Chandrasekhar (1942), and the wave-mode theory by Zonn {\&} Rudnicki
(1957). See also (Lindblad 1962; Contopoulos 1972; Toomre 1977, 1996; Pasha
2000).}$^{,}$\footnote{ In Lindblad's bar-mode theory as it had
progressed by the early 1950s (Lindblad {\&} Langebartel 1953), three
factors serve for the spiral formation. The first is the tendency for the
formation of the rings, one at the galaxy center and one (or several) more
in the distance, the bar occupying the inter-ring region. The second factor
is the development of two diametrically opposed zones of enhanced
density (see Fig.2). The third one is the increased centrifugal (radial)
motion in these zones. If the bar-forming processes affect the galaxy
kinematics but weakly, then the motions of distant material lag behind that
of the main galactic body, and as the existing radial motions make the outer ring deform and break up, it forms the main spiral arms (I and II in Fig.3). Also, the effects of the bar wave show that
material at the bar `tips' has some extra rotation, so that, helped by the
radial motions, it forms the inner spiral arms (VI in Fig.3). If the
galactic angular momentum is above some certain level, the density wave can
give no bar, and the deviations from axial symmetry it causes produce the
appearance of ordinary spiral structure.}

\vfill\eject

\begin{figure}
\centerline{\epsfxsize=0.7\columnwidth\epsfbox{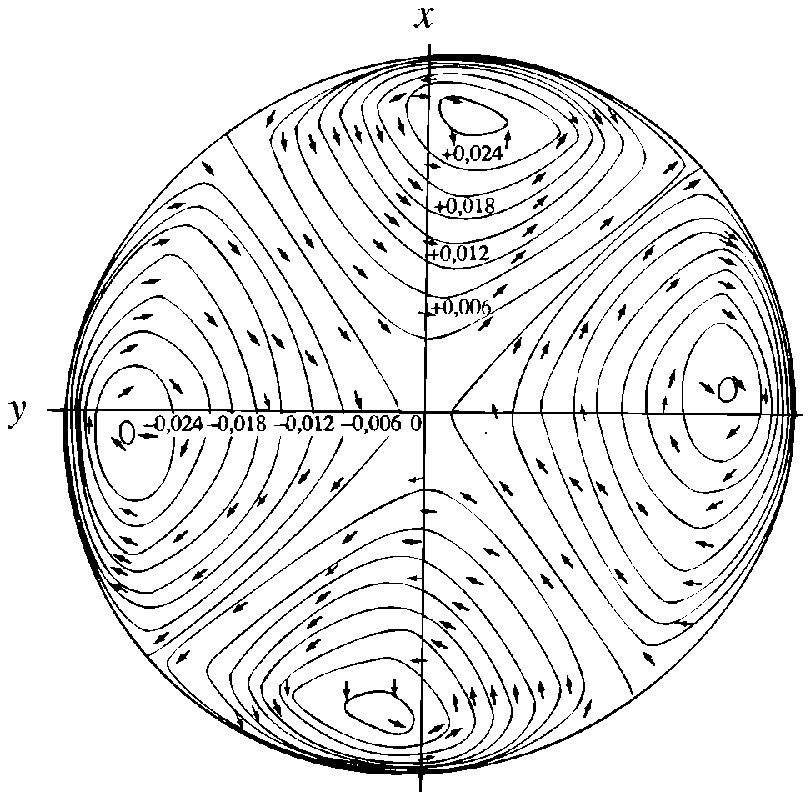}}
\caption{\footnotesize \textit{The m = 2 wave mode in Lindblad's
bar-spiral density wave theory}. Two wave maxima and minima are placed along the $x$ and $y$ axes, respectfully. These bisymmetrically located maxima and some extra concentration at a galaxy center are to explain the bar phenomenon. The arrows show systematic noncircular motions. (The figure is reproduced from Lindblad {\&}
Langebartel 1953)}
\bigskip
\centerline{\epsfxsize=0.6\columnwidth\epsfbox{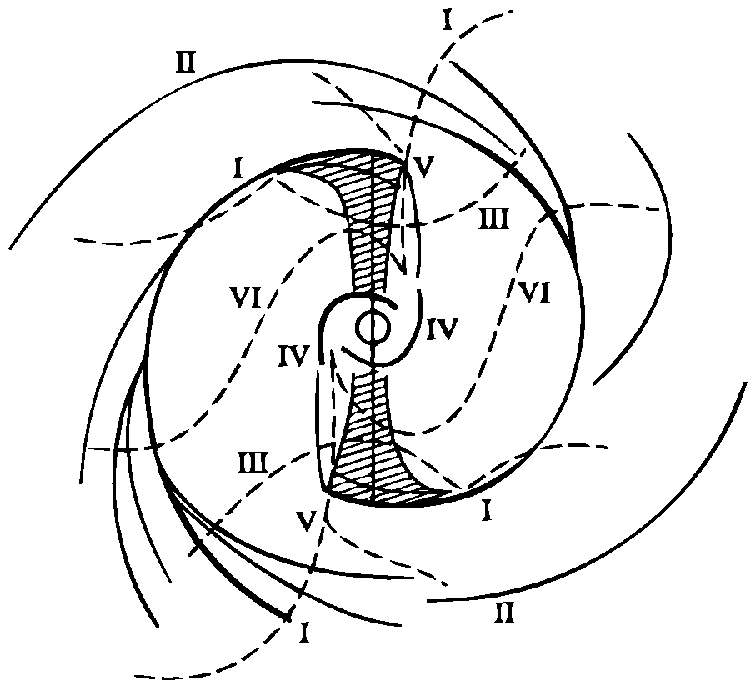}}
\caption{\footnotesize \textit{The formation of spiral structure as envisaged in
Lindblad's bar-spiral density wave theory}. (The figure is
reproduced from Lindblad {\&} Langebartel 1953)}
\end{figure}

\subsection*{1.2 Gas and dust}
\addcontentsline{toc}{subsection}{1.2 \it Gas and dust}

\bigskip

{\footnotesize \begin{list}{}{\leftmargin4cm}
\item The difficulty of cosmogonical theories lies in the interconnection of the
facts.
\begin{flushright}
\textit{Weizsacker 1951, p.165}
\end{flushright}

\item Where a few years ago we seemed to be up against a blank wall of
discouragement, we are now in an era of rapidly developing research.
\begin{flushright}
\textit{Bok {\&} Bok 1957, p.244}
\end{flushright}
\end{list}}

\bigskip

\noindent Stellar dynamics of the 1940s - early 1950s was essentially the theory of a
stationary galaxy arranged by the regular forces (see Ogorodnikov 1958) and
the theory of quasi-stationary systems open to slow relaxation processes
(Ambartsumian 1938; Chandrasekhar 1942, 1943). Together, they provided a
basis serving well for getting certain practical dividends but still of
little use for conceiving the underlying dynamical problems.

\begin{quotation}{\footnotesize
\noindent ``While these methods have contributed substantially toward the
clarification of the peculiarly characteristic aspects of stellar dynamics,
an impartial survey of the ground already traversed suggests that we are
perhaps still very far from having constructed an adequate theoretical
framework in which the physical problems can be discussed satisfactorily. In
any case we can expect that the near future will see the initiation of
further methods of attack on the problems of stellar dynamics''
(Chandrasekhar 1942, p. vii-viii).\footnote{ ``I remember very vividly the
atmosphere in the 50's in stellar dynamics. On the one hand, we had the
most general solutions of Liouville's equation by Chandrasekhar. But it was
realized that the self-consistent problem required also the solution of
Poisson's equation, which was very difficult in general. Thus people were
discouraged.'' (\textit{Contopoulos})} \par}
\end{quotation}

The envisaged future did not happen to lie as immediately near, however. The
theoretical thought kept on whirling around the idea of galaxies
evolutionarily tracking over the Hubble diagram, one way or the other, and
that opened in quite a few attempts at a synthesis of the available strict
knowledge about gravitating figures in a softer (then bulkier) spirit of
cosmogonical inclusion.\footnote{ See, e.g., the ``Critical review of
cosmogonical theories prevailing in West Europe and America'' by Schatzman
(1954). It would be some fuller with an addendum on a theory developed in
1955-56, now in the Soviet Union, by Ogorodnikov. Finding that the works by
Lindblad and Chandrasekhar on collisionless dynamics ``really bar the way to
studying the laws of evolution of stellar systems'', he suggested a ``more
promising'' -- ``synthetic'' -- hydrodynamical method with elements of
statistical mechanics (Ogorodnikov 1958, p.20, 22), and with this he proved
theorems on uniform rotation and nearly constant density for ``dynamically
determinable'' systems, at their ``most probable phase distribution''. This
enabled Ogorodnikov to start his supposed evolutionary sequence with the
`needle-shaped' galaxies, or strongly elongated ellipsoids in rotation about
their shortest axis. Such needles are secularly unstable, above all at their
long-axis extremities from where ``the stars are detached in two winding
arms'' giving the picture of a typical barred spiral galaxy. Material
released during this gradual bar destruction feeds a spherical halo, while
inside the bar a violent process of low-velocity-dispersion star formation
starts, and these emerging Population I stars uniformly fill the new
equilibrium figure -- a thin disk-like Maclaurin spheroid. The remaining
diffuse material of the bar (needle) winds up and, being still `frozen' in
the disk, forms spiral arms. Due to irregular forces, Population I and II
stars get mixed, because of which the spiral galaxy cannot be in
equilibrium: its disk dies out through dissipation, and a nuclear remainder
drives up an eventual elliptical galaxy (Ogorodnikov 1958, p.29). \par
As well illustrative appears Weizsacker's theory of galaxies and stars built
on a concept of supersonic turbulent motion in the original gaseous mass,
the one picturing a general ``evolutionary trend as far as it does not depend
on the special conditions by which galaxies, intragalactic clouds, stars, planets,
etc., are distinguished''. The theorist understands the rapid flattening of that
gaseous mass (in about one period of rotation) as due to the decay of its original
turbulence, and he reduces its further evolution to some secular changes followed
by a slow loss of the axial rotation of the galactic systems. In this way, galaxies
of the type of the Magellanic Clouds or the M31 companions are to be obviously
younger than the universe, and ``elliptic galaxies are in a final stage which no
longer shows the sort of evolution we consider''. ``Thus the large galaxies like
our own can be as old as the universe, without having yet reached their final
stage'', the spiral structure being their ``most conspicuous semiregular
pattern''. Weizsacker's judgment on it is twofold. He finds himself in a
position to ``try to understand spiral structure as a hydrodynamical effect
[\ldots ] produced by nonuniform rotation'', noticing that any local
formation -- ``cloud formed by the turbulence'' -- will then be distorted
into a segment of a spiral. On the other hand, he admits that ``the
abundance of systems with just two spiral arms is probably caused not by
turbulence but by gravitation'', which is in fair correlation with the
presence of a bar. The bar is understood as an elongated equilibrium figure
of rotation similar to Jacobi's liquid ellipsoids; it ``can be kinematically
stable only if the system rotates uniformly'', i.e. in inner galactic
regions. But just a little way out, the shearing effect of differential
rotation comes into play, in order ``not to destroy the `bar' entirely but
to distort it strongly'', giving it some spiral contours (Weizsacker 1951,
p.176-179).} Accordingly, non-stationary -- dynamical -- problems of
deformation of the systems and of density disturbances in them seemed
difficult and therefore premature, while stationary problems were held as
``natural and necessary'' at that preliminary point, for ``it is hard to
imagine that at all stages the evolution of stellar systems has the
violently catastrophic character'' (Ogorodnikov 1958, p.13).\footnote{
Zwicky reflected on the `cooperative' effects in gravitating systems (both
in stars and galaxy clusters) since the mid-1930s, and he believed that
whereas the nuclei of spiral galaxies had already reached their equilibrium
the spiral arms and interarm regions were still ``transitory
configurations'' (Zwicky 1957, p.214). He thus did not treat the spiral
structure from the natural, for collective phenomena, viewpoint of
oscillations and waves in equilibrium media.} In this illumination,
Lindblad's theory of unstable bar-modes was typically deemed extravagant and
unacceptable (Lebedinski 1954, p. 31).

\begin{quotation}{\footnotesize
\noindent ``Such theories cannot yet help the progress of cosmogony, since
uncertainty in them still prevails validity'' (Schatzman 1954, p.279). \par}
\end{quotation}

The delicacy of this sort of expert judgment -- let alone its other virtues
-- reflected clearly that it was the issue of gas and dust that became a
common focus of galaxy astronomy despite its stellar past.\footnote{ ``Why
do the spirals always show the combination of a disk and a central
spheroidal system? It must reflect the original density distribution in gas.
[\ldots ] Can we imagine that at some era in the past, the central
spheroidal system of low rotation and the disk with very fast rotation
actually resembled the equilibrium figure of the gas? One should really look
into these things'' (Baade 1963, p.17).\par ``The origin of the spiral
systems is an unsolved problem as yet. Doubtless the interstellar material
plays a major part in it. Therefore the methods [of stellar dynamics \ldots
] seem to be insufficient for a solution'' (Kurth 1957, p.146).} By the
1950s, Baade discovered in M31 many hundreds of emission nebulosities (HII
regions), having concluded that ``they are strung out like pearls along the
arms'' (Baade 1963, p.63). Gas and dust, he stated, are also distributed in
this galaxy highly unevenly, grouping in its spiral arms.\footnote{ This was
inferred from the lack of reddening of globular clusters in M31, one half of
which lie behind the galaxy disk because of their spherical distribution. As
Baade wrote (1963, p.70), initially one did not believe in this finding,
since the gas layer in our own Galaxy was still held to be uniform.}
Besides, no one already doubted the youth of high-luminosity stars since
they were ascertained to still form in abundance, e.g. in the Orion nebula.
The sheer weight of these individually weak facts convinced many workers
that

\begin{quotation}{\footnotesize
\noindent ``the primary phenomenon in the spiral structure is the dust and gas, and
that we could forget about the vain attempts at explaining spiral structure
by particle dynamics. It must be understood in terms of gas dynamics and
magnetic fields'' (Baade 1963, p.67).\footnote{ Baade has usually been
quoted from his posthumous monograph (Baade 1963). It reproduces his 1958
lectures that vividly transmit the mid-century atmosphere in extragalactic
astronomy. Many investigators of the time claimed to have agreed with Baade
on the basic role of gas in the spiral arrangement (e.g., Weizsacker 1951,
p.178).} \par}
\end{quotation}

The lion's share of these discoveries was made possible due to the 200-inch
Palomar reflector put into operation in 1949, although from 1951 onwards the
interstellar gas was unprecedentedly attacked also by the 21-cm-line
methods. Dutch radio astronomers presented ``one of the truly historic
diagrams of Milky Way research'' (Bok {\&} Bok 1957, p.244) -- a detailed
map of atomic hydrogen distribution (Hulst et al 1954).\footnote{ In 1958
this map was completed with the spiral fragments observed from Australia
(Oort et al 1958).} It displayed extended fragments of tightly-wrapped
spiral arms which in the solar vicinity matched `local arms' in Sagittarius,
Orion and Perseus.\footnote{ They were inferred in 1951 from data on the
distribution of O-B associations and HII regions (Morgan et al 1952; see
Gingerich 1985).} Gas kinematics routinely analyzed, a synthesized rotation
curve of the Galaxy was pictured (Kwee et al 1953), and the ``primary task
for the next few years'' was claimed to get improved radio equipment
``capable of tracing with precision the spiral structure of our Galaxy''.

\begin{quotation}{\footnotesize
\noindent ``While there is always room for theorizing, the emphasis must first of all
be on careful observation and unbiased analysis of observations"''(Bok {\&}
Bok 1957, p.248). \par}
\end{quotation}

The new empirical facts -- the tightly wrapped, nearly ring-like arms of the
Milky-Way spiral, the concentration in them of Population I objects, the
\textit{general} shearing character of rotation -- were a surprise to Lindblad. He could not
neglect them. But they demanded another, more fitting dynamical theory, and
Lindblad put aside (but did not deny\footnote{ Via such shifts of opinion,
Lindblad found himself on the way towards ``a more definite theory''
(Lindblad 1962b, p.148). There he might well be judged (Toomre 1977, p.439)
as if even having finally conceded that his old leading-arm models were
``not reconcilable with modern evidence'' (Lindblad 1962b, p.146). Yet he
blamed that on some other ``early gravitational theories which interpret
spiral structure as due to orbital motions of stars starting from a small
nucleus'' (Lindblad 1962b, p.146).}) his business with unstable circular
orbits and wave bar-modes. This step was largely favored by first numerical
experiments in galaxy dynamics performed in 1955-60 by his son P.O. Lindblad
with the big electronic computing machine installed in Stockholm (Lindblad
{\&} Lindblad 1958; P.O. Lindblad 1962). Those experiments showed the
\textit{trailing} -- not the leading -- spiral arms, the ones supported by fresh data on both
the form of the Milky-Way spiral and the space orientation of many galaxies
(Vaucouleurs 1958), and, after all, the ones put into orbit way back by
Hubble (1943) in the framework of his working hypothesis that galactic
spirals \textit{always} trail.\footnote{ Having completed by the 1930s his theory of asymptotic leading spirals, Lindblad (1934) turned to the empirical
component of the problem of the `sense of rotation' of spiral arms. The
difficulty was with determining the near and the far sides of a galaxy, as
this might be made no other than by way of speculation on the asymmetry of
dust absorption along the minor axis of the visible image. There were at the
time no reliable data on interstellar dust properties. To Lindblad's way of
thinking, a stronger absorption was felt by a farther side (thought also to
show sprinkles of dust veins in the bulge region), which maintained leading
arms. After a categorical objection by Hubble (1943), he scrutinized the
subject anew in his fundamental work with Brahde (Lindblad {\&} Brahde 1946)
followed by a succession of smaller articles during a decade or so. To
criticize Lindblad for his leading-arm orientation was a commonplace. One
agreed with him (and, evidently, with Hubble) in that the sense of spiral
winding must be the same for all galaxies, which demanded only one good
example of a nearly edge-on galaxy that might be clearly judged on both its
spiral form and nearer side. Vaucouleurs (1958) gave such an example as got
a high-quality long-exposure photograph of NGC 7331 taken with the 200-inch
reflector. It favored Hubble's camp. Lindblad must have reserved objections
on how the spiral form was to be inferred from that crucial case (he and his
collaborators Elvius and Jensen had been studying this galaxy
photometrically in several papers from 1941 to 1959, and he gave a rather
incomplete summary on the topic in Lindblad 1962a), but for the absolute
majority of astronomers the empirical component of the sense-of-winding
problem was no longer acute.}

\bigskip

\subsection*{1.3 Winds of change}
\addcontentsline{toc}{subsection}{1.3 \it Winds of change}

\bigskip

{\footnotesize \begin{list}{}{\leftmargin4cm}
\item The spiral structure is nothing more than a tracer element contained in a
fairly uniform disk of material [\ldots ] This is probably related to the
magnetic field in the disk.

\begin{flushright}
\textit{G. R. Burbidge 1962, p.295}
\end{flushright}

\item As far as I am aware, no single problem, not even a stability problem, has
been solved in a differentially rotating self-gravitating medium. Even
without magnetic fields, and even linearizing the equations, it is very hard
to make progress.

\begin{flushright}
\textit{Prendergast 1962, p.318}
\end{flushright}

\item With our observations we have reached a point where we are simply unable to draw any definite conclusion, unless the theory helps us. I hope some day
there will be action, because otherwise we are lost.

\begin{flushright}
\textit{Baade 1963, p.266}
\end{flushright}
\end{list}}

\bigskip

\noindent The post-war success in galaxy research gave priority to the empirical
approach. By the late 1950s, it formed two flanks of evolutionary studies,
morphological and quantitative. The first one, due mostly to the Palomar sky
survey, called for elaborate classifications, catalogs and atlases of
galaxies (Zwicky 1957; Morgan {\&} Mayall 1957, de Vaucouleurs 1959;
Vorontsov-Velyaminov 1959; Sandage 1961); the second exploited matters
concerning stellar evolution and empirical data on individual galactic
objects. As regards the theoretical approach, it too branched under the new
conditions and its subject was now treated in distinct frames of physical,
chemical and dynamical evolution.

\bigskip

On this dynamical side, the one to our present interest, true lodestars
started shining by the 1960s. One of them was lit by the linear stability
theory as applied to long-range force systems; denied so far, mostly by
human inertia, its methods eventually penetrated into the galaxy
dynamics.\footnote{ ``I cannot agree that plasma physics methods penetrated
in astronomy in the 50's. Of course these developments helped each other,
mainly in the 60's, but this is natural. I think that in the 50's progress
was sporadic, due to the insight of only a few people, but later many people
followed the first pioneers''. (\textit{Contopoulos})} Chandrasekhar (1953, p. 667) formulated
the problem as follows:

\begin{quotation}{\footnotesize
\noindent ``When we know that an object has existed in nearly the same state for a long time we generally infer that it is stable; and by this we mean that
there is something in its construction and in its constitution which enables
it to withstand small perturbations to which any system in Nature must be
subject. [\ldots ] Thus when we are confronted with a novel object -- and
most astronomical objects are novel -- a study of its stability may provide
a basis for a first comprehension''. \par}
\end{quotation}

To him, however, it was a matter of pure intellectual interest, above all.
``For an applied mathematician, Chandrasekhar explained, problems of
stability present a particular attraction: by their very nature, these
problems lead to linear equations and linear equations are always more
pleasant to deal with than nonlinear ones'' (Chandrasekhar 1953, p.667).\footnote{ Particularly, this was the line in which the unified theory
of ellipsoidal equilibrium figures was being developed later (Chandrasekhar
1969). ``There was criticism by astronomers of Chandrasekhar's work on the
classical ellipsoids because of its remoteness from the current needs of
astronomy. Chandra's interest (and my own as well) was indeed motivated by
non-astronomical considerations. What we found was a development by some of
the great mathematicians of the 19th and early 20th century that had largely
been forgotten, and in some mathematical respects was left incomplete. Chandra felt strongly that his work should, on general intellectual grounds,
be completed. If that completion should have application in astronomy, so
much the better, but that was not the motivation. His critics in astronomy
were offended because he was not doing astronomy. Chandra, however, was more
devoted to science (or his view of it) than to astronomy, and did not feel
obligated to work on problems which were chosen for him by astronomers''.
(\textit{Lebovitz})} In so thinking, he turned to most general, technically transparent models. One of such was Jeans' infinite homogeneous medium asked about whether the classical stability criterion $k^{2}c^{2} - 4\pi G\rho > 0$ and the critical fragmentation scale $\lambda _{J} = (\pi c^{2} / G\rho )^{1 /
2}$ remain unchanged if the medium is involved in uniform rotation ($Я$ and
$\rho $ are sound speed and material volume density; $k$, $\omega $ and
$\lambda = 2\pi / k$ -- wave number, frequency and length; $G - $gravity
constant).\footnote{ ``I do remember that at the time I wrote the paper, the
spiral structure of the galaxies was not even remotely in my mind. Besides
my paper was concerned with the Jeans instability of a gaseous medium and
not to a system of stars\ldots However, I am quite willing to believe that
the basic ideas were included in earlier papers by Lindblad''.
(\textit{Chandrasekhar})} The answer came positive, with the one exception for perturbations
propagating in the direction just at right angles to the rotation axis, when
Coriolis force co-governs wave dynamics and modifies the dispersion relation
into

\begin{equation}
\omega ^{2} = 4\Omega ^{2} - 4\pi G\rho + k^{2}c^{2}
\end{equation}

\noindent showing that any rotation with $\Omega > (\pi G\rho
)^{1 / 2}$ entirely prevents the system from decay.

\bigskip

Safronov (1960a,b), interested in protoplanetary cloud dynamics as a part of
his solar-system cosmogony, examined a more realistic model -- a
differentially rotating gas layer stratified along the rotation
axis.\footnote{ Ledoux (1951), interested in the formation of planets from a
primordial cloud, seems to have been the first to consider the stability of
flat gravitating systems. He, as well as Kuiper who had turned him to this
problem, suspected a change in the critical Jeans scale, realizing that an
assumed cloud mass of about 10{\%} that of the Sun would be enough for the
cloud to act significantly on itself in the plane of symmetry. Ledoux found
that for small adiabatic disturbances to the equilibrium state of an
isothermal non-rotating layer Jeans' criterion remains unaltered if $\rho $
is taken to be half the density value at $z = 0$. This did give only a
correction to the clumping scale, which was of order $2\pi $ times the
thickness. Fricke (1954) combined the efforts by Ledoux (1951) and
Chandrasekhar (1953), yet he too could not escape certain arbitrary
assumptions. And Bel {\&} Schatzman (1958), having returned to
Chandrasekhar's model, let it rotate differentially -- in violation of the
equilibrium conditions, though.} A short-wave analysis led him to a relation

\begin{equation}
\omega ^{2} = \kappa ^{2} - 4\pi G\rho \cdot f(k,h) + k^{2}c^{2}
\end{equation}

\bigskip

\noindent that basically differed from Eqn (2) in its modified
gravity term depending on both wavenumber and the layer's
thickness $h$. The correction factor$ f(k$,$h)$ evaluated,
Safronov found -- quite in Jeans' spirit -- that rotating flat
systems lose stability and must break up into rings as soon as
their equilibrium volume density gets above some critical value.

\bigskip

In that same 1960, first results were supplied by collisionless collective
dynamics, concerning the simplest, spherical systems.\footnote{ Vlasov, a
renowned plasma physicist, contributed to galaxy dynamics as well, via his
article (Vlasov 1959) that had a special section ``Spiral structure as a
problem of the mathematical theory of branching of solutions of nonlinear
problems''. Through the collisionless Boltzmann and Poisson equations, he examined the equilibrium
of an immovable plane-parallel slab, re-derived its density profile $\rho
(z)\sim $ sech$^{2}(z$/$h)$, and `disturbed' eigenvalues of the equilibrium
solution, wishing to establish the character of ``infinitely close figures of
equilibrium''. His new solutions turned out ``ribbed'', or spatially
periodic, with the ``exfoliation period'' being close to 3 kpc and
corresponding to the scale of ``stellar condensations observed by Oort''.
Despite some technical flaws (e.g., his basically smooth function $\rho
(z)$ played as stepped one in integrations), Vlasov's conclusion about
possible ``ribbed'' static equilibria in the tested slab was formally
correct. Still, surprisingly (at least in retrospect), he gave no stability
discussion, already practicable in contemporary plasma physics and very fitting as it would be for his galactic model.} Antonov (1960) found for them the now
classical ``stability criterion, rather complicated though'', and
Lynden-Bell (1960a) discovered a peculiar feature of their equilibrium
states -- the ability of collisionless spheres to rotate.\footnote{ ``This
is in contradiction to Jeans' result, but is obtained by using his method
correctly and following the consequences'' (Lynden-Bell 1960a, p.204).}

\bigskip

Another lodestar for dynamical studies was the evidence provided by a bulk
of higher-precision rotation curves obtained for spiral galaxies in the late
1950s by Burbidges and Prendergast. At long last, their general rotation was
ascertained to be strongly differential. This fact, stripped now of all
surmise, seriously warned astronomers that they were in the presence of a
real problem of the \textit{persistence} of spiral structure.

\begin{quotation}{\footnotesize
\noindent ``There appears to have been some feeling in recent years that individual
spiral arms are long-lived features in a galaxy. [\ldots ] However [\ldots ]
we shall show that the form of the rotation-curves for spirals will insure
that the spiral form will be completely distorted in a time short compared
with the age of a galaxy'' (Prendergast {\&} Burbidge 1960, p.244). \par}
\end{quotation}

The quantitative estimates did show that the data on M31, M81, NGC 5055
``and probably all similar spiral galaxies'' were in conflict with ``certain
apparently reasonable assumptions'' -- namely, at least with one out of the
following three: (a) only circular velocities are present in galaxy disks,
(b) these velocities are independent in time, (c) material which is
originally in a spiral arm remains in that arm (Prendergast {\&} Burbidge
1960, p.244, 246).

\bigskip

The `urgent problem' of the persistence of spiral forms was taken up by
Oort. Speaking at a 1961 conference at Princeton of ``every structural
irregularity'' in a galaxy as being ``likely to be drawn out into a part of
a spiral'', he called for another phenomenon to turn to and conceive:

\begin{quotation}{\footnotesize
\noindent ``We must consider a spiral structure extending over a whole galaxy, from
the nucleus to its outermost part, and consisting of two arms starting from
diametrically opposite points. Although this structure is often hopelessly
irregular and broken up, the general form of the large-scale phenomenon can
be recognized in many nebulae'' (Oort 1962, p.234). \par}
\end{quotation}

Oort suggested ``three ways out of this difficulty'', one of which was that
``the arms could retain their present spiral shapes if matter were
constantly being added to their inner edges, while the outer edges would
constantly lose matter'' (Oort 1962, p.237-8). This possibility was given
an eager discussion at the conference (Oort 1962, p.243).

\bigskip

Yet one more lodestar for galaxy dynamics was lit in the 1950s by numerical
computer methods. They first served the calculating of three-dimensional
star orbits; Contopoulos (1958, 1962) then stated their non-ergodicity and
posed anew the problem of a third integral of motion. P.O. Lindblad, as we
saw, turned the same Stockholm computer to studying the galaxy dynamics in
terms of an $N-$body problem (Lindblad {\&} Lindblad 1958; P.O. Lindblad 1962).

\bigskip
\bigskip

\subsection*{1.4 Dispersion orbits}
\addcontentsline{toc}{subsection}{1.4 \it Dispersion orbits}

\bigskip

{\footnotesize \begin{list}{}{\leftmargin4cm}
\item Most remarkably after that fine beginning [in 1925-27], it took Lindblad not
three further months or years, but three whole decades, to connect this
implied epicyclic frequency $\kappa $ and the ordinary angular speed of
rotation $\Omega $ into the kinematic wave speeds like $\Omega \pm \kappa /
m$, which we very much associate with him nowadays, especially when
muttering phrases like `Lindblad resonances'.

\begin{flushright}
\textit{Toomre 1996, p.2-3}
\end{flushright}
\end{list}}

\bigskip

\noindent These fresh winds did not catch Lindblad unawares. The importance of
differential rotation was already conceived by him from radio observations
(Kwee et al 1954; Schmidt 1956), and he even noticed -- for the Galaxy and,
later, for M31 (van de Hulst et al 1957) and M81 (Munch 1959) -- the curious
empirical near-constancy of a combination

\bigskip

\begin{equation}
\Omega _{2} = \Omega (r) - {{\kappa (r)} \mathord{\left/ {\vphantom {{\kappa
(r)} {2}}} \right. \kern-\nulldelimiterspace} {2}} \cong const.
\end{equation}

\bigskip

\noindent And the dynamical stability problems were
\textit{always} comprised by his spiral theories. Already from
1938 on, dispersion relations of type (3) surfaced in his
evolving papers, growing more and more complicated by way of
various gradient-term inclusions for a tentatively better
description of the crucial -- unstable -- bar-mode (see Genkin
{\&} Pasha 1982).\footnote{ Lindblad's dispersion relation in its
simplest form (Lindblad 1938) was rather similar to Safronov's
relation (3), both showed the same terms, but, as Lindblad was
focused on global modes and Safronov dealt with short-wave radial
oscillations only, their treatment of the correcting factor in
gravity term was technically different. Still, ``Lindblad,
despite all his words, never quite seemed to relate those
formulas to any \textit{spiral} structures, and [\ldots ] only
applied them literally to non-spiral or bar-like disturbances''.
(\textit{Toomre})}

\bigskip

However, the idea of applying the collective-dynamical methods to shearing
stellar galaxies hardly ever impressed Lindblad. He must have felt (Lindblad
1959) the limits of his hydrodynamical approach (long-wave solutions at
differential rotation were unattainable analytically, while, on the
short-wave side, the whole approach failed for want of an equation of
state), not having yet a means of solving kinetic equations. Also, Lindblad
perhaps doubted the very possibility of steady modes in shearing galaxies.
Either way, the empirical relation (4) that he himself had stated inspired
him the most. With it as a centerpiece he started a new, ``more definite
theory of the development of spiral structure'' (Lindblad 1962b, p.148),
one he called the \textit{dispersion orbit} theory (Lindblad 1956, 1961). It was imbued, intuitively,
with a hope that gas and Population I stars ``are somehow aggregated on
their own into a few such orbits in each galaxy -- almost like some vastly
expanded meteor streams'' (Toomre 1996, p.3).

\bigskip

Lindblad described epicyclic stellar oscillations in a reference system
rotating with angular velocity $\Omega _{n} = \Omega - {{\kappa}
\mathord{\left/ {\vphantom {{\kappa}  {n}}} \right.
\kern-\nulldelimiterspace} {n}}$, $n = {{d\kappa}  \mathord{\left/
{\vphantom {{d\kappa}  {d\Omega}} } \right. \kern-\nulldelimiterspace}
{d\Omega}} $, and he imagined a star's radial displacement $\xi $ to depend
on its azimuth $\theta $ as $\cos n(\theta - \theta _{0} )$, $\theta _{0} $
being apocentric longitude. The simplest forms of orbits occurred for
integer $n$'s, the case of $n = 2$ satisfying the empirical condition (4).
For this case, ``the most general form of an ellipsoidal distribution with
vertex deviation'' was obtained (Lindblad 1962b, p.152), with which
Lindblad sought to calculate the total gravitational potential and, by
extracting its averaged (over time and angle) part, to treat the remainder
as a contribution to the perturbing force. He Fourier-decomposed this force
and retained the $m = 1$, 2 harmonics to analyze disturbances
to a ring of radius $r$ composed of small equal-mass particles. Like Maxwell
(1859) in his similar Saturn ring problem,\footnote{ Maxwell's problem was
on disturbances of $N$ equal-mass particles placed at the vertices of an
$N$-sided regular polygon and rotating in equilibrium around a fixed central
body.} Lindblad obtained four basic modes for each $m$. Two of them described
nearly frozen, practically co-rotating with material, disturbances to the
ring density. Two others -- ``deformation waves'' -- ran with speeds $\Omega
\pm \kappa / m$, the minus sign being for the slower mode. It was, at $m$ = 2,
``essentially this slowly advancing kinematic wave [\ldots] composed of many
separate but judiciously-phased orbiting test particles'' (Toomre 1977, p.441) that Lindblad meant by his dispersion orbit $\xi (\theta )$. The fact
that its angular velocity was independent of radius, $\Omega _{p} (r) =
\Omega _{2} = $\textit{const} (with an observational accuracy of the condition (4)),
implied a stationary state for all test rings, i.e. over the entire radial
span where this condition was well obeyed.

\begin{quotation}{\footnotesize
\noindent ``This fact greatly intrigued Lindblad -- who did not
need to be told that strict constancy [of $\Omega _{p} (r)$]
would banish wrapping-up worries or that the nicest spirals tend
to have two arms. Yet astonishingly, that is about as far as he
ever got. [...] It never occurred very explicitly to [him ...] to
combine already those `orbits' into any long-lived spiral
\textit{patterns}'' (Toomre 1977, p.442). \par}
\end{quotation}

\bigskip

\subsection*{1.5 Circulation theory of quasi-stationary spirals}
\addcontentsline{toc}{subsection}{1.5 \it Circulation theory of
    quasi-stationary spirals}

\bigskip

{\footnotesize \begin{list}{}{\leftmargin4cm} \item The suggestion that the patterns are density waves is old and was first
explored by Bertil Lindblad. His emphasis was mainly on kinematics and less
on collective effects on a large scale, though many of the kinematical
effects he discovered can still be seen in the collective modes.
\begin{flushright}
\textit{Kalnajs 1971, p.275}
\end{flushright}

\item His details were unconvincing, but no one can accuse him of missing the big
picture.

\begin{flushright}
\textit{Toomre 1996, p.3}
\end{flushright}
\end{list}}

\bigskip

\noindent P.O. Lindblad's experiments with flat galaxies were planned to clarify the dispersion-orbit theory. They started with a plane system of several annular
formations arranged by $N \cong $ 200 mutually attracting points, and the
development of ``small deviations in shape and density of a bisymmetrical
nature'' (Lindblad 1963, p.3), applied to one of the rings, was studied. Two
waves propagating along it were shown to rise first, one running slightly
faster and the other slower than unperturbed particles, thus invoking a pair
of corotation resonances, one on each side from the ring. These induced a
leading spiral; soon it rearranged into a trailing one and smeared out
almost completely, but some trailing arms then re-appeared, owing evidently
to a small oval structure retained at the center. This led P.O. Lindblad to
propose that galactic spirals may involve a \textit{quasi-periodic} phenomenon of trailing-arm
formation, breakup and re-formation.\footnote{ ``I was delighted to see them
[P.O. Lindblad's results] as evidence as to how much one could do already
then (!) by way of interesting numerical studies with some hundreds of
particles -- in that sense his work was very inspiring. Yet [\ldots ] it
also struck me that his study really dealt with not much more than the
transient breakup of inherently unstable configurations of some 4 or 5
artificially introduced rings of material'' that imitated ``a revolving disk
-- one which [\ldots ] should be fiercely unstable if begun just as cold.
[\ldots ] But, again, as a sample of what could already be done, P.O.
Lindblad's work was indeed like a breath of fresh air''. (\textit{Toomre})}

\bigskip

B. Lindblad, however, got captivated by another view of these results. He
even lost of his earlier dispersion-orbit enthusiasm and turned in 1961-62
to a concept ``\textit{On the possibility of a quasi-stationary spiral structure in galaxies}'' (Lindblad 1963) in the presence of differential
rotation.\footnote{ Lebedinski was another one who in his cosmogony of
galaxies and stars admitted -- still earlier -- ``the dynamical possibility
of the formation of quasi-stable spiral arms rotating with a constant
angular velocity for all the spiral'' (Lebedinski 1954, p.30). Yet since
Jeans' 1920s that idea, as such, did not sound as a novel dynamical motive.
It got a really new sounding only when the fact of global galactic shearing
was finally conceived.}

\begin{quotation}{\footnotesize
\noindent ``The morphological age of spiral galaxies as estimated [\ldots ] from
considerations of the evolutionary process connected with star formation
from gaseous matter ranges between $10^{9}$ and $10^{10}$ years. In
consequence it is natural to assume that the typical spiral structure is not
an ephemeral phenomenon in the systems but has a certain steadiness in time
[\ldots and] to investigate how far gravitational forces alone can explain a
spiral structure of a fair degree of permanence'' (Lindblad 1964, p.103). \par}
\end{quotation}

To begin with, Lindblad introduced an axisymmetric flat stellar system in
differential rotation and, echoing the $N-$body pictures, imposed on it an
initial \textit{trailing} spiral pattern formed by some extra amount of stars. His
calculations of the effect upon a nearby test star from such a spiral arm
showed that, as it sheared, the star approached it and fell in, having no
other chance to leave it than making slight epicyclic oscillations. Such an
assimilation of material in just one galactic turn or so worked well against
shearing deformation of spiral arms, through their exchange in angular
momentum with stars attracted. As the result, the pattern's angular speed
became the same all over, meaning its quasi-stationarity. Now two
dynamically different regions arose in the system, an inner region with
stars moving faster than the spiral, and an outer one, tuned oppositely;
they were divided by a corotation region, where the material orbits at
nearly the same rate as the pattern.

\bigskip

For a true stationary pattern not only its permanence in shape was needed,
but also a balance of the stars' travel in and out of the arms. The latter
was secured in Lindblad's eyes by his \textit{circulation} theory (Lindblad 1963, 1964)
developed in the framework of a trailing two-armed spiral model, each arm
making one full convolution (or a bit more), comparably inside and outside
corotation (Fig.4). Actually, each arm ended where, according to analytical
estimates, its stars were effectively attracted by the next-to-last arm
(outside corotation) and fell in it ``in a shower of orbits''. The
assimilated stars kept moving slower than the spiral, thus having an
along-arm ascent until a repeated flow down. Inside corotation (the region
of much less interest to Lindblad), the circulation was set up as well, but
in the opposite direction: stars captured by spiral arms got drawn down
along them until sucked upward by the next-to-innermost spiral convolution.

\begin{figure}
\centerline{\epsfxsize=0.9\columnwidth\epsfbox{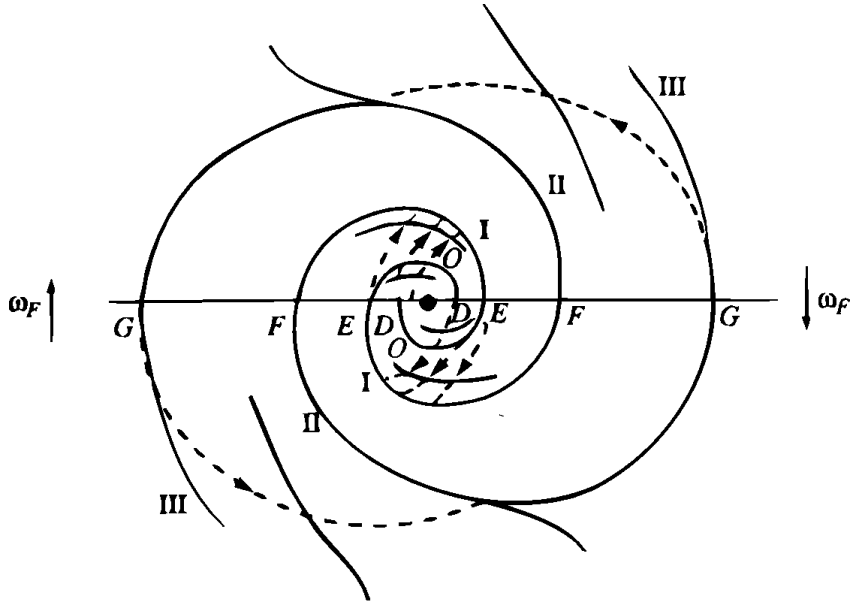}}
\caption{\footnotesize \textit{Circulation of material in a
galaxy having a quasi-stationary spiral structure.} The general
rotation is clockwise, points $F$ mark the corotation radius. See
the text for more details. (The figure is reproduced from
Lindblad 1964)}
\end{figure}

\bigskip

This circulation theory was nothing but a sketch by 1964. Well treating
quasi-steady spirals as a density \textit{wave}, it gave no desired quantitative results
regarding pattern speeds, arm pitch angles, interarm spacings, or the like.
It also failed to explain \textit{dynamically} the preference for trailing arms -- although the
dispersion-orbit theory had honestly done no better. It is regrettable that
Lindblad, who died in 1965, did not have the time to complete this last work
he had started, and only ``left behind a long handwritten unfinished
manuscript that in great mathematical detail studies the gravitational
effects of spiral arms in his circulation pattern'' ($P$.$O$. \textit{Lindblad}).

\bigskip

\begin{center}
* * *
\end{center}

\bigskip

\noindent The original spiral theories by Bertil Lindblad passed into oblivion. Among
the causes for the passage were the feeble empirical base of the 1920s-40s,
the frightening bulk of mathematics and scant help from the first computers
even during the 1950s, a constant flux of changes in Lindblad's latest
inferences and the rather opaque prose of his abundant articles,\footnote{
``It has not been possible to do justice to all phases of Lindblad's
researches'', Chandrasekhar `complained' already in 1942, but nonetheless he
gave a ``more or less complete bibliography'' including 25 Lindblad's
writings on the spiral problem (Chandrasekhar 1942). ``The flow of his
publications can be understood if one realizes that he thought in the form
of a paper. When attacking a problem he started writing the paper at once''.
(\textit{P.O. Lindblad})\par}  and above all a lack of quantatively checkable predictions. Yes,
one can readily agree that

\begin{quotation}{\footnotesize
\noindent ``all problems that in later developments turned out to be important in the
theory of spiral structure had, in one way or another, already been touched
upon or even studied by Lindblad'' (Dekker 1975, p.18) \par}
\end{quotation}

\noindent
as well as that

\begin{quotation}{\footnotesize
\noindent ``such complex collective dynamics was perhaps too hard for anyone, no
matter how talented, in those mid-20$^{th}$-century decades before
computers, plasma physics, or any inkling of massive halos'' (Toomre 1996,
p.3), \par}
\end{quotation}

\noindent
but also true is that all of the spiral undertakings by Lindblad, however
ingenious and farsighted they may appear to have been in retrospect, got
sunk ingloriously in the silence of time.

\bigskip

An interesting question is: \textit{why}? Why did it come to be that the true master of
theory and observation had long been surprisingly close to but never quite
at the point of recognition -- opened in the 1960s to a pleiad of fresh
theorists -- that spiral structure is mainly a collective wave phenomenon in
shearing galaxies? One can only suppose that Lindblad did not reach, let
alone exploit, such wave-mechanical ideas partly because they were not in
the air yet, but perhaps mainly because he was impeded by his life-long
emphases on the orbits of individual particles. \textit{All} his efforts on galaxy
dynamics were fed by the stellar-epicycle concept, the pearl of his
scientific youth. This set the trend for Lindblad's theories, and whenever
some such orbital attack fell short of its destination, he did not get on
with searching for totally different ways of continuing, but instead renewed
his attack time and again under his old epicyclic-orbit colors.

\bigskip
\bigskip

\section*{II. ON A NEW WAVE CREST}
\addcontentsline{toc}{section}{II. On a new wave crest}

\bigskip

{\footnotesize \begin{list}{}{\leftmargin4cm}
\item During a time when it was fashionable to `explain' the maintenance of spiral
structure by magnetic fields, Lindblad persisted in the belief that
gravitation was the dominant factor, and now we have come full circle back
to this view.
\begin{flushright}
\textit{E.M. Burbidge 1971, p.266}
\end{flushright}
\end{list}}

\bigskip

\subsection*{2.1 Regenerative spirals by Lynden-Bell}
\addcontentsline{toc}{subsection}{2.1 \it Regenerative spirals by Lynden-Bell}

\bigskip

{\footnotesize \begin{list}{}{\leftmargin4cm}
\item We deduce that our galaxy is likely to have had spiral arms for most of its
lifetime and that as old arms coil up so new uncoiled arms must start to
form from their corpses. The problem of describing such a mechanisms we call
the regeneration problem.
\begin{flushright}
\textit{Lynden-Bell 1960b}
\end{flushright}
\end{list}}

\bigskip

\noindent In 1960 Lynden-Bell presented at the University of Cambridge his PhD thesis
``Stellar and Galactic Dynamics'' (Lynden-Bell 1960b)\footnote{ Leon Mestel
was his advisor.}  considering some general aspects of stellar-dynamical and
ergodic theories. Its separate part ``Cosmogonical gas dynamics'' was on the
spiral problem. It stated, echoing the stress of the day, that ``the arms
are primarily the seat of gas and dust'' (so that the lenticular galaxies,
deprived of them, ``can no longer give birth to a spiral structure''). It
found the cosmogonical approach the most convenient -- in case of full
denial from Jeans' classic scheme as inoperable in the presence of
differential rotation.

\begin{quotation}{\footnotesize
\noindent ``It seems impossible that the protogalactic gas was uniformly rotating when
the stars formed. It seems more likely that as the primordial gas broke up
into condensations [protogalaxies] each fluid element tended to preserve its
angular momentum about the centre of the local condensation. The equilibrium
reached is then one in which centrifugal force nearly balances gravity and
the pressure is mainly important in preventing the system from becoming very
flat.'' \par}
\end{quotation}

Lynden-Bell analyzed realistic equilibrium configurations of a frictionless
gas system and derived ``an energy principle which should provide a powerful
means of determining the equilibria on a computer''. Any such configuration,
when achieved by the system, is exposed to a slow secular evolution that
``will not be determined by shrinkage due to the radiation of energy as in
Jeans case, but by the transfer of angular momentum due to friction''
neglected in the equilibrium derivations. The system ``must: i) concentrate
its angular momentum into a very small fraction of its total mass, and ii)
leave the remainder a more concentrated uniformly rotating or pressure
supported body. This is borne out by observation on both the scale of the
solar system and that of the galaxy. [\ldots ] We should thus expect a
uniformly rotating central condensation surrounded by a differentially
rotating disc'' (Lynden-Bell 1960b).

\bigskip

It is with such an evolved disk of gas that Lynden-Bell linked his spiral
considerations. In shearing deformation -- a point-blank menace to `any
structural irregularity' -- he, unlike many workers of the day, saw not an
antagonist to the persistence of spiral arms, but a factor of their cyclic
regeneration created through gravitational instability of the gaseous
subsystem in a combined star-gas galactic disk (the stellar component being
liable for gas equilibrium rather than for any collective dynamics). In such
a setting, the problem needed a global stability analysis of a system in
differential rotation, which technically was not feasible. That is why for
want of the better Lynden-Bell employed the methods that had served Fricke
(1954) with his $\Omega = const$ model; this led to a necessary and
sufficient condition of Jeans' stability, ${{\Omega ^{2}} \mathord{\left/
{\vphantom {{\Omega ^{2}} {\pi G\rho _{0} > 2 / 3}}} \right.
\kern-\nulldelimiterspace} {\pi G\rho _{0} > 2 / 3}}$ (cf. Sect. 1.3), and instructed the growth rate for unstable stages
to be $\gamma \le 2\Omega $. An $m$ = 2 mode at $k \cong 1 / 3$ kpc$^{ - 1}$
was found the most important, it fell down towards the disk edge and center,
being long-wave and therefore fast-growing. This was in substance Lindblad's
bar mode, one specified by a pair of condensations placed oppositely at $r
\cong 9$ kpc from the center. Before density had grown by a factor $e$, rotation
turned the system through 180$^{0}$ (at $\gamma = 2\Omega )$. But as this
passed, effects of shear (excluded from the strict stability analysis) just
wound the ``azimuthally independent structure'' round the galaxy, at least
once. This meant a grave radial-wavelength reduction, which was expected to
be a cause for slowing down the growth rate as effectively as to turn off
instability altogether. In this event, the spiral arms would expand back
``to form the sheet from which we started'', and the whole process might
then recur. However, a more careful analysis confirmed the dependence of
$\gamma $ on $k$ only ``for systems very close to stability''. This would be
``far too sensitive to give the great variety of spirals'' and could not
apply ``for any part of the observed spiral arms''. The regeneration theory
proposed, Lynden-Bell (1960b) concluded, was ``therefore untenable''.

\bigskip

But as it turned out later, this pessimism was rather excessive, since it
became clear eventually that there was a good deal of wisdom even in such
regenerative thoughts. This, however, is not how things developed
immediately, because, as we will see in the forthcoming section, the old
idea of steady spiral modes was about to gain a new and important burst of
enthusiasm.

\bigskip

\subsection*{2.2 MIT enthusiasm}
\addcontentsline{toc}{subsection}{2.2 \it  MIT enthusiasm}

\bigskip

Chia Ch'iao Lin was not an astronomer. Since the pre-war time, he had been
studying fluid flows. By the 1960s, he had had over 60 publications, a
monograph on hydrodynamic stability (Lin 1955), a world recognition of an
applied science expert, and a solid reputation at the department of
mathematics in the Massachusetts Institute of Technology (MIT) where he
worked since 1947. But he did feel a continual interest in astronomy, being
admired with strict analytical papers by Chandrasekhar, with M.
Schwarzschild' work on stellar structure, with Zwicky's morphological
method. In 1961 this side interest became Lin's life-long vitality. That
spring, on visit in Princeton,\footnote{ Stromgren invited him for
discussions on stellar structure (\textit{Lin}), largely in relation to his fresh
interest in hydrodynamics of liquid helium (Lin 1959).} he attended the
aforementioned conference on interstellar matter and, having become familiar
with the developments in galaxy research, he got captured by the problem of
the persistent spiral structure.\footnote{ In his early spiral papers, Lin
often quoted Oort's statement reproduced in Sect. 1.3.}

\bigskip

Back in MIT, Lin conveyed his galactic enthusiasm to his young colleagues
Hunter and Toomre.\footnote{ At that time, the department of mathematics in
MIT was vigorously enlarging its applied side. Hunter and Toomre were hired
there in 1960, just after they had got their PhD degrees in fluid dynamics
in England. Initially, they hoped to collaborate with Backus (\textit{Hunter}; \textit{Toomre}), a
recognized leader in geomagnetic problems, but as he left MIT that year
already, they two ``soon caught some of Lin's fever for problems in the
dynamics of galaxies''. ``Almost at the moment I first met him in fall 1960
I was struck with his breadth of scientific interests, his really excellent
spoken English, [\ldots ] and his genuinely gracious manner of dealing with
other people''. (\textit{Toomre})} For quick acquaintance with current periodicals, a
`reading group' was formed;\footnote{ ``[We] were all becoming interested in
astrophysical problems together. We read Martin Schwarzschild's book on
stellar structure together''. (\textit{Hunter})} a ``friendly back-and-forth atmosphere''
(\textit{Toomre}) warmed open discussions and working visits of Woltjer and Lust, organized
by Lin;\footnote{ ``It was a real pleasure to have such a thoughtful and
articulate theoretical astrophysicist as Woltjer so close to chat with about
this thing or that. [\ldots ] It was from his informal lectures that summer
that I learned for the first time not only how Dutch and Australian radio
astronomers working in parallel had more or less mapped the spiral arms of
this Galaxy from the velocity maps, but also how astonishingly thin -- and
yet curiously bent -- is our layer of 21-cm gas''. (\textit{Toomre})} Lebovitz was hired in
the department.\footnote{ ``I had just received my PhD [working with
Chandrasekhar], I wished to pursue applied mathematics, and I had received
an offer of an instructorship from one of the best applied-mathematics
departments in the country. Lin's motive I can only speculate on. He was
interested in moving in the direction of astronomy and of the
spiral-structure problem and perhaps figured I would be a useful
participant. If this is the case, I suppose my stay at MIT may have been
somewhat disappointing to him because I spent all of it in close
collaboration with Chandrasekhar on a quite different set of problems''.
(\textit{Lebovitz})} In 1962, Shu arrived there for doing his undergraduate course work under
Lin's guidance,\footnote{ ``I began work with C.C. Lin in summer 1962 as an
undergraduate research assistant and continued through the fall and spring
1963, on the topic of spiral structure in galaxies as my undergraduate
thesis project in physics at MIT [\ldots ] I knew Lin from even earlier
because he is a close friend of my father''. (\textit{Shu})} and Hunter with Toomre,
their instructorship finished, left MIT, one back for Cambridge, UK, the
other for Princeton; their first papers appeared in 1963.

\bigskip

Hunter and Toomre made their debut in galaxy dynamics on a vital problem
already posed but yet unanswered very basically (Kuzmin 1956; Burbidge et al
1959): How to connect the empirical rotation curves of galaxies with their
equilibrium mass distribution? Toomre (1963) set forth a general
mathematical method, and for a razor-thin disk model he derived a series of
solutions well known nowadays as Toomre's models of $n^{th}$ order (Binney
{\&} Tremaine 1987, p.44).\footnote{ Toomre's model 1 reproduced the result
by Kuzmin (1956) then unknown to Toomre (Binney {\&} Tremaine 1987, p.43).}
Hunter (1963) used a distinct thin-disk approximation and found another
series of exact solutions. The simplest there was the case of uniform
rotation and surface density $\mu _{0} (r) \propto (1 - {{r^{2}}
\mathord{\left/ {\vphantom {{r^{2}} {R^{2})^{1 / 2}}}} \right.
\kern-\nulldelimiterspace} {R^{2})^{1 / 2}}}$. For it only was the
analytical study of equilibrium stability possible, and Hunter did it ``using
only pencil, paper, and Legendre polynomials'' (Toomre 1977, p.464). This
cold disk proved \textit{unstable} for a wide span of axisymmetric and non-axisymmetric
oscillation modes.\footnote{ The stability of differentially rotating cold
disks Hunter studied in his subsequent paper (Hunter 1965).} These papers by
Toomre and Hunter had paved the way for further works on kinematical models
and global dynamics of flat stellar systems.

\bigskip
\bigskip

\subsection*{2.3 Gravitational stability of flat systems}
\addcontentsline{toc}{subsection}{2.3 \it Gravitational stability of flat systems}

\bigskip

{\footnotesize \begin{list}{}{\leftmargin4cm}
\item Lin asked [Woltjer in 1961]: What are the circumstances that would be needed
for either one or both of the stellar and interstellar parts of a supposedly
smooth galactic disk to remain gravitationally stable against all large
scale disturbances?
\begin{flushright}
\textit{Toomre 1964, p.1217}
\end{flushright}

\item The importance of collective effects in our Galaxy was first clearly pointed
out by Toomre (1964). He showed that in the disk the stellar motions are
sufficiently coherent to make it almost vulnerable to collapse. He also
pointed out that the scale on which this would occur is quite large.
\begin{flushright}
\textit{Kalnajs 1971, p.275}
\end{flushright}
\end{list}}

\bigskip

\noindent As we have seen, Safronov already raised the question of gravitational
instability in flat rotating systems, aiming at the breakup of a
protoplanetary cloud into detached rings. Toomre, interested in basically
smoother objects like galaxies, turned in 1961 to a rather close, although
opposite in accent, topic, and by the summer of 1963 he prepared an article
``On the gravitational stability of a disk of stars'' (Toomre 1964,
hereinafter T64).

\bigskip

The paper started with the general presentation of the problem as it was
then seen.

\begin{quotation}{\footnotesize
\noindent ``The well-known instabilities of those Maclaurin spheroids whose rotational
flattening exceeds a certain fairly moderate value suggest that the other
sufficiently flattened, rotating, and self-gravitating systems might in some
sense likewise be unstable. At any rate, these instabilities have been often
cited as a likely reason why one does not observe elliptical galaxies
exceeding a certain degree of oblateness. It is only when we turn to
consider what are now thought to be the distributions of all but the
youngest stars in the disks of the ordinary (as opposed to the barred)
spiral galaxies that this classical result suggests a serious dilemma: How
is it conceivable, in spite of these or analogous instabilities, that so
much of the fainter stellar matter within such galaxies -- and certainly the
S0 galaxies -- should today appear distributed relatively evenly over disks
with something like a ten-to-one flattening?'' (T64, p.1217) \par}
\end{quotation}

\begin{figure}
\centerline{\epsfxsize=0.6\columnwidth\epsfbox{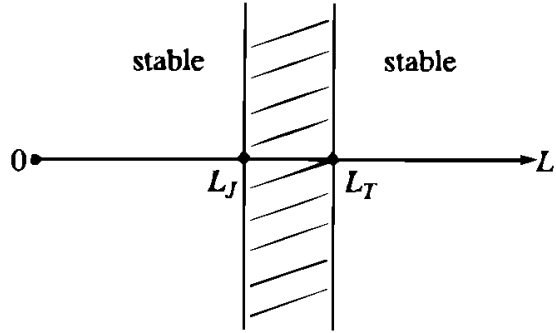}}
\caption{\footnotesize \textit{Characteristic scales in a gravitating disk}. A cold rotating disk is stable for radial disturbances on the scales $L >
L_{T} $, a non-rotating hot disk is stable of scales $L
< L_{J} $, a hot rotating disk is stable on both scales. As the velocity dispersion becomes of the order of the circular velocity, one obtains full axisymmetric stability.}
\end{figure}

The detailed study of the problem was preceded by a primary, qualitative
stability estimate.

\bigskip

A rotating thin cold disk, in an approximate equilibrium between gravity and
centrifugal forces acting on each mass element, is prevented from general
contraction, still not from fragmentation. Small-size clumpings arise
everywhere in such a disk, and then collapse, their gravity taking excess
over rotation. But if larger-sized, they do not go as these two factors
counteract each other. The demarcation length scale $L_{T} $ proves plain
co-measurable with the disk radius $R$. Thus the cold model, for all
specifications it may have, is clearly unstable.\footnote{ Toomre got this
estimate by the fall of 1961 and was struck with the fact that nothing had
ever been said on the thing just shocking with its as simply derivable
inference that cold disks be prone to violent instabilities. (\textit{Toomre})} The part
played by random motions is best visualized with an immovable sheet model.
There instability is avoided if stars (other mass elements), having an rms
velocity $Я$, cross a clumping zone in a time not exceeding that needed for an
$e-$fold amplitude growth as registered in the cold case. Hence the largest yet
ungrowing disturbance is found on an $L_{J} \approx c^{2}$/$G\mu _{0} $
scale, which is essentially the Jeans stability criterion. Now, letting the
sheet rotate, one sees the two characteristic scales, $L_{T} $ and $L_{J} $,
be present (Fig.5). $L_{J} $ gets closer to $L_{T} $ for higher velocity
dispersions, until they coincide at $c$'s as high -- in the order of magnitude
-- as the rotational velocity, thus meaning full stabilization against this
sort of disturbances.

\bigskip

The strict analysis of \textit{axisymmetric} disturbances to a razor-thin disk, performed in T64,
supported these rough estimates. In the cold case, it led to a local
dispersion relation
\begin{equation*}
\omega ^{2} \; = \;
\kappa ^{2} - 2\pi G\mu _{0} {\left| {k} \right|} \eqno(5a)
\end{equation*}

\noindent or

\begin{equation*}
\nu ^{2} = 1 - {\left| {k} \right|} / k_{T}  \eqno(5b)
\end{equation*}
\addtocounter{equation}{1}

\noindent linking the wave frequency in units of $\kappa $, $\nu =
\omega / \kappa $, with a critical wavenumber
\begin{equation}
k_{T} = \kappa ^{2} / 2\pi G\mu _{0} ,
\end{equation}

\noindent the one to determine the shortest wavelength $\lambda
_{T} \equiv 2\pi / k_{T} $ of ungrowing ($\nu ^{2} \ge $ 0)
disturbances (Fig.6).\footnote{ Analyzing axisymmetric
disturbances to a flattened rotating cloud, Safronov (1960a,b)
did not solve the Poisson equation. He was guided by the notice
that short radial waves find adequate the cylindric approximation
for a torus (ring). But the cylinder is the sum of `rods', or
elementary cylinders whose individual gravity is given by a
simple formula, so that the business is just to integrate in
infinite limits the elementary contributions over longitudal and
transversal variables $x $and $z$. There Safronov was not
perfect, however. His gently stratified cloud turned a stiff
2$h-$thick plate as he took his introduced density function $\rho
_{0} (z)$ out of integration over $z$. His subsequent integration
over $x $was in an interval of $\pm \lambda / 4$; that, he
argued, ensured a predominant contribution to the perturbed force
(which is qualitatively true). Had he integrated in infinite
limits, and first -- most trivially -- over $x$, the gravity term
in his Eqn (3) would have become $ - 2\pi Gk\int {\rho _{0} (z /
h)e^{ - k{\left| {z} \right|}}dz} $, and with the exponential
factor serving as a thickness correction he would have accurately
managed with any density profile -- and, most obviously, would
have found that in the zero-thickness limit that factor
simplifies to unity, the integral just gives the surface density
$\mu _{0} $, so that the gravity term converts into --$2\pi G\mu
_{0} k$, the form in which it was presented soon by Toomre (1964)
in frames of `regular' methods of the potential theory.}  The
hot-disk analysis detected the minimum radial velocity dispersion
at which the system is still resistant against \textit{all}
axisymmetric disturbances (Fig.7):\footnote{ To solve the Vlasov
kinetic equation, Toomre used the characteristics method that for
some three-dimensional purposes had already served Lynden-Bell
(1962), who in his turn cited the original source (Bernstein
1958) where that method had genuinely helped with the general
disperion relation for the mathematically similar problem with a
Maxwellian plasma in a magnetic field.}$^{,} $\footnote{ Because
of a technical error in Toomre's analysis, this minimum value was
initially overestimated by 20{\%}. Not so little if one considers
that the difference in $c_{r,\min} $ for star and gas disk models
(the latter case admits a \textit{much} simpler analysis) reaches
7{\%} only. It is this ``substantial error'' which was detected
in 1963 by Kalnajs (cf. Sect. 2.4), as reported frankly in T64
(p.1233).}

\begin{equation}
c_{r,\min}  = {\frac{{3.36G\mu _{0}}} {{\kappa}} }.
\end{equation}

\bigskip

\noindent The real-to-minimum velocity-dispersion ratio

\begin{equation}
Q = {{c_{r}}  \mathord{\left/ {\vphantom {{c_{r}}  {c_{r,\min}} } } \right.
\kern-\nulldelimiterspace} {c_{r,\min}} }
 \ge 1
\end{equation}

\bigskip

\noindent thus got a local disk-stability parameter.\footnote{
Formally, the `$Q-$parameter' (8) was introduced by Julian and
Toomre (1966).}$^{, }$\footnote{ This quantitative analysis
refines the above view of disk stabilization as it shows via Eqs
(6) and (7) that locally the result is attained already once
$L_{J} / L_{T} = $ (3.36/2$\pi )^{2} \cong $ 0.286 (0.25 in a gas
disk).} In a \textit{marginally} stable state $Q = 1$,
disturbances of $\lambda _{0} \cong 0.55\lambda _{T} $ proved
most unpliant and barely suppressible. Our solar neighborhood
would have such a $\lambda _{0} \cong 5 - 8$ kpc, but if some $Q
\cong 1 - 1.5$ were not preferred empirically, implying a certain
stability reserve. Of course, ``it was as yet impossible to rule
out instabilities altogether'', but should any actually be
present, they would not do with scales responding to the
challenging 2-kpc spacings, as these ``must almost certainly be
judged as stable''. This ``is important as an argument against
any suggestion that the existing spiral structure in this Galaxy
might be the result of collective stellar
\textit{instabilities}'' of the sort considered (T64, p.1236).

\begin{figure}
\centerline{\epsfxsize=0.9\columnwidth\epsfbox{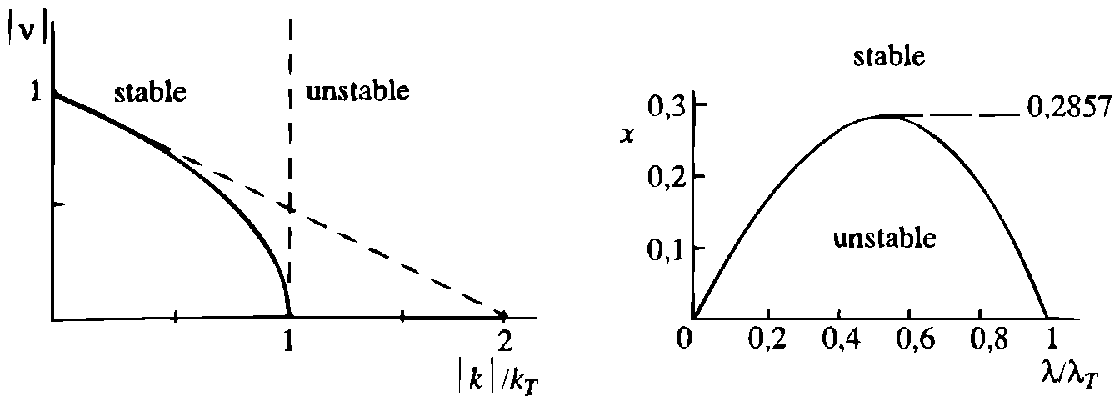}}
\caption{\footnotesize (left) \textit{The dispersion relation curve for radial
oscillations and tightly wrapped spiral waves in a cold disk}.}
\caption{\footnotesize (right) \textit{The hot disk neutral stability curve}. The disk is stable for all those the radial disturbances for which the parameter $x =
k^{2}c_{r} ^{2} / \kappa ^{2}$ exceeds $x_{cr} = 0.2857.$
This critical value determines the minimum velocity dispersion (7) sufficient to secure the axisymmetric disk stability. (The figure is reproduced from Toomre 1964)}
\end{figure}

\bigskip
Still, the linear theory developed could not lay claim to very much. So it
did not elucidate the cause of stellar disk heating, it even could not show
any definitely what was to become with primary condensations appearing in a
tentatively cold disk in one or two revolutions already. ``It must not be
presumed that such initial clumpings would necessarily have led to the
formation of any \textit{permanent} irregularities'', Toomre noticed. ``On the contrary, it
seems much more likely that the bulk of the stars involved in any given
(generally non-axisymmetric) instability [\ldots ] would eventually have
dispersed themselves upon emerging from the opposite sides of the  aggregation and upon experiencing the shearing effect of differential
rotation''.


\begin{quotation}{\footnotesize
\noindent ``It follows that an initially unstable disk of stars should probably have
undergone not just one but several successive generations of instabilities,
after each of which the system would have been left somewhat less unstable
than it was previously. In particular, it seems likely that before very many
rotation periods had elapsed, the disk would have approached a new
equilibrium state that was again fairly regular and quite possibly
axisymmetric, but in which the random velocities at the various radii had
become -- and would henceforth remain -- about equal to the minimum values
needed for complete stability'' (T64, p.1237).\footnote{ Asked to reminisce
on how he had originally understood those dispersion velocities ``about
equal'' to the needed minimum in the new equilibrium state -- on whether or
not this was a factual suggestion of marginal stability of our stellar disk,
or some extra amount was yet permitted for its stability -- Toomre has
responded: ``It is hard for me to reconstruct from this vantage point what
exactly I meant or hoped by that statement. Probably I was mostly just
trying to rationalize the surprising fact which I had then unearthed that
the minimum theoretically needed $c_{r,\min}  $ and the observed amounts
seemed to agree so well within their considerable uncertainties, meaning
within a factor of 1.5 or thereabouts, rather than some 2 or 3 or 4 [\ldots]
From about 1966 onwards, I was surely of the opinion that any $Q_{} $ less
than about 1.5 here was highly suspect, if not downright ludicrous, because
of fierce heating of cooler disks by their embedded gas complexes. But that
came a little later. In 1964 my views were no doubt more permissive toward
$Q$ = 1.0''. (\textit{Toomre})} \par}
\end{quotation}

Besides, since the total gravitational energy of the disk would have had to
be the same during its evolution (the virial theorem), ``the said
redistribution of stars could not simply have consisted of an overall
contraction, but would have had to entail a contraction perhaps in the inner
parts of the disk jointly with a net expansion of the outer portions'' (T64,
p.1237) -- as it was already seen by Lynden-Bell (1960b) from the
gas-dynamical viewpoint.

\bigskip

As regards \textit{non-axisymmetric} disturbances, it was pointed out in T64 that because of the
specific action of the Coriolis force those are restrained even more
effectively than radial disturbances, thus requiring no addition for
$c_{r,\min}  $. However, Toomre remarked, a question that his discussion
left ``completely unanswered'' was ``to what extent a similar amount of
random motion [$Q = 1$] might affect the character of the most extensive
non-axisymmetric disturbances, in particular those which ought to determine
whether or not a given disk might prefer to develop into a barlike
structure'' (T64, p.1235).\footnote{ Real progress in the study of this
problem first came half a decade later.}

\bigskip
\bigskip

\subsection*{2.4 Kalnajs' search for spiral modes}
\addcontentsline{toc}{subsection}{2.4 \it Kalnajs' search for spiral modes}

\bigskip

{\footnotesize \begin{list}{}{\leftmargin4cm}
\item One can draw a parallel between the attempts to talk about galactic
evolution at the present time and the attempts to understand stellar
evolution before the sources of energy in the stars were understood.
\begin{flushright}
\textit{G. R. Burbidge 1962, p.291}
\end{flushright}

\item The study of stellar systems, such as our own galaxy, is not limited by a
lack of understanding of the underlying principles, but rather by the
difficulty of solving the differential equations which govern the time
evolution of the system.
\begin{flushright}
\textit{Kalnajs 1962, p.i}
\end{flushright}
\end{list}}

\bigskip

\noindent Agris Kalnajs began his undergraduate studies in Electrical Engineering at
MIT in 1955. As a good student, he participated in a special course which
emphasized physics and mathematics, and provided summer employment in the
Microwave Research Lab at Raytheon, making measurements for computer
modeling of magnetrons. There he learned about such things as electron
motions in crossed electric and magnetic field, waves carrying positive and
negative energies, modes, coupled modes, parametric amplification. All this
proved to be really useful in a quite different field when he arrived in
1959 in the astronomy department at Harvard University and got involved in
galaxy dynamics.\footnote{ ``It was probably David Layzer's course in
classical dynamics which steered me towards stellar dynamics. I rather liked
David's approach: he strived for elegance. He put a lot of thought in his
lectures''. (\textit{Kalnajs})}

\bigskip

In the fall of 1961 Kalnajs made a research examination on ``Stellar
kinematics'' (Kalnajs 1962).\footnote{ As this was only an unpublished
internal document, its outline below is mainly to illustrate how Kalnajs was
then progressing.} The task was to calculate self-consistent radial
oscillations in a rotating stellar disk as a tentative explanation for the
`local' arms in our Galaxy. Their short spacing $L \le 3$ kpc justified the
small-scale analysis in the plane of a homogeneous thin sheet. Kalnajs
solved the Vlasov and Poisson equations as an initial-value problem and
obtained an equation for the radial oscillations and a dispersion relation
which was formally correct.\footnote{ Following Landau's method correctly
describing small oscillations in homogeneous electrostatic plasma, an
arbitrary disturbance is initially imposed on the stellar sheet and its
evolution is traced out. With time, the dependence on the initial conditions
dies away, and the result is provided by the integrand poles whose
expression -- the dispersion relation -- connects the established wave
parameters.} As he was interested in short waves, he made an asymptotic
evaluation of the integral expression, and in the process left out ``a
factor $2\pi $ or something of that order'' (\textit{Kalnajs}). This and the reduced disk
response at the short waves ($\lambda \sim 1$kpc) made him conclude that
$\omega \cong \kappa $, because the self-gravity effects became ``too small
to be interesting'' (\textit{Kalnajs}): all the solutions oscillated and were traveling
waves that, in passing, ``tend to gather up the low dispersion objects such
as gas'' (Kalnajs 1962, p. ii). As a plausible ``arm-like density wave''
generator, an oval-shaped body at the Galaxy center was mentioned.

\bigskip

The error in this asymptotic evaluation was uncovered in the summer of 1963
when Kalnajs and Toomre finally got together, compared and crosschecked
their notes, and detected each other's technical errors. Kalnajs looked anew
at his radial-oscillation theory and re-evaluated the dispersion relation,
this time into the form in which it entered his thesis (Kalnajs
1965).\footnote{ ``Strictly speaking, I was the first to write down the
dispersion relation. But that is not the important thing. What is more
important is who made the best use of that equation. And here it was Toomre,
who used it to discuss the stability of the Galactic disk -- a distinctly
more fundamental topic than the subject of my Research Examination. [\ldots
] By the time we got together in 1963, that is probably the way we
understood our respective contributions''. (\textit{Kalnajs})} In modern notation
-- whose convenience and clarity we owe undoubtedly to Lin -- and without the
uninteresting stellar disk thickness correction going through that original
1961-63 analysis,\footnote{ The thickness corrections were worth considering
for wavelengths as short as 1.5 kpc as they reduced the radial force by a
factor of 2 or 3, but for $\lambda \cong 6$ kpc the reduction was some
20{\%}-30{\%} at most.} it is

\begin{equation}
\nu ^{2} = 1 - {{{\left| {k} \right|}} \mathord{\left/ {\vphantom {{{\left|
{k} \right|}} {k_{T} \cdot F_{\nu}  (x),}}} \right.
\kern-\nulldelimiterspace} {k_{T} \cdot F_{\nu}  (x),}}
\end{equation}

\noindent where

\begin{equation}
F_{\nu}  (x) = 2(1 - \nu ^{2}){\frac{{e^{ - x}}}{{x}}}{\sum\limits_{n =
1}^{\infty}  {{\frac{{I_{n} (x)}}{{1 - {{\nu ^{2}} \mathord{\left/
{\vphantom {{\nu ^{2}} {n^{2}}}} \right. \kern-\nulldelimiterspace}
{n^{2}}}}}}}} ,
\quad
x \equiv {{k^{2}c_{r} ^{2}} \mathord{\left/ {\vphantom {{k^{2}c_{r} ^{2}}
{\kappa ^{2}}}} \right. \kern-\nulldelimiterspace} {\kappa ^{2}}},
\end{equation}

\bigskip

\noindent is Kalnajs' version of a factor to account for the role
played by random motions of stars. There is no such play in the
limit $x = 0$, relation (9) then reduces to Toomre's cold-disk
result (5) that shows the gravity term proportional to the
wavenumber and growing without bound. Now random motions arrest
this growth: the total contribution of gravity only reaches a
maximum at $x_{0} \cong 1$, still giving rise to instability
($\nu ^{2} < 0)$ if large enough, and for $x > > 1$ it becomes
small. In the solar neighborhood that value of $x_{0} $ points to
a radial wavelength $\lambda _{0} \cong 6$ kpc, the one concluded
by Toomre from his neutral stability analysis. Its
commensurability with the radial size of the Galactic disk makes
the local theory somewhat suspect.

\begin{quotation}{\footnotesize
\noindent ``When I wrote my Research Examination I was under the impression that the spacing between the spiral arms was about 1.5 kpc. After Toomre and I got
together, it became clear to me that the 1.5 kpc waves/fluctuations were not
the important modes of the Galaxy. [\ldots ] Also by the fall of 1963 I had
obtained my own copy of Danver's thesis (thanks to my uncle who was at Lund
University). Danver had measured the spiral patterns and came up with a
typical pitch angle of $16^{\circ}.6$. This implies scales even larger than
6 kpc. [\ldots ] By this time Alar had published his disk models, and I
could use them to estimate the scales at which these disks were most
responsive, and they convinced me that a WKBJ approach [see Sect. 3.1] was
too crude [\ldots ] and that -- unlike plasma -- galaxies were too
inhomogeneous. [...] So the future was `global modes and integral
equations'.'' (\textit{Kalnajs}) \par}
\end{quotation}

Once he realized this fact, Kalnajs lost interest in the local theories,
which were good for the stable small-scale solutions, and turned to \textit{global} modes
as the correct approach to the oscillation problem. In the fall of 1963 he
presented to his thesis committee at Harvard ``An outline of a thesis on the
topic `Spiral structure in galaxies' '' (Kalnajs 1963), summarizing his
ideas for a new theory of steady spiral waves. Because this document has
been almost unknown, a long quotation from it appears to be quite
appropriate.\footnote{ ``I do not recall exactly when I first learned that
Lin was also interested in spiral density waves (it was probably a talk he
gave at MIT), but at that stage our relations were most cordial and I also
felt that my understanding of this topic was more thorough than his. So
having produced a written document, I am pretty sure that I would have found
it difficult \textit{not} to boast about my achievements'' (\textit{Kalnajs}). ``A written document''
there refers to the ``Outline'' which at least Toomre received from Kalnajs
in November 1963.}

\begin{quotation}{\footnotesize
\noindent ``A feature peculiar to highly flattened stellar systems is the appearance
of spiral markings, called arms. These features are most prominently
displayed by the gaseous component of the galaxy and the young hot stars
which excite the gas. However, the density fluctuations can still be seen in
the stellar component, appearing much fainter, but also more regular.

\bigskip

The division of the galaxy into two components, gaseous and stellar, appears
natural when one considers the dynamical behavior of these two subsystems.
The gaseous component is partly ionized and is therefore subject to magnetic
as well as gravitational forces, and has a very uneven distribution in the
galactic plane. The stellar system is quite regular, its dynamics being
governed by the long-range gravitational forces arising from the galaxy as a
whole; the density of stars is sufficiently low that binary encounters
between stars may be ignored. The stellar component, which is the more
massive, cannot support density fluctuations on a scale much smaller that
the mean deviation of the stars from a circular orbit (or the scale of the
peculiar motions). The gas, on the other hand, would support smaller-scale
fluctuations -- at least in the absence of magnetic effects. The fact that
observed spiral arms are not much narrower than the smallest scale that the
stars will tolerate suggests that stars must participate actively in the
spiral patterns.

\bigskip

There is a fundamental difficulty, however, in the assumption that spiral
arms are entirely stellar: if an arm can exist and does not grow in time,
then its mirror image is also a possible configuration. This follows from
the time-reversibility of the equations of motion combined with their
invariance under spatial inversion. Thus the leading or trailing character
cannot be decided on the basis of a linearized theory if we insist on
permanency of the spiral markings. The observations indicate, however, that
nature in fact prefers trailing spiral arms. Thus a plausible theory of
spiral structure must include both the stars and the gas.

\bigskip

I regard the galaxy as consisting of two components, gas and stars, coupled
by gravitational forces. The stars provide the large scale organization and
the gas discriminates between leading and trailing arms. ([\textit{Footnote in the original text}]: The stellar
system can be thought of as a resonator, and the gas would then be the
driver which excites certain of the normal modes.) If the coupling is not
too strong, one may at first consider the two subsystems separately, and
afterwards allow for their interaction. Unfortunately, one cannot evaluate
the magnitude of the coupling without calculating the normal modes of the
two subsystems. For the gaseous component, only the crudest type of analysis
is possible at present, since one should include non-linear terms in the
equations governing the gas motion in order to be realistic. The stellar
component, on the other hand, is sufficiently smooth that a linearized
theory should apply, and the problem of determining the normal modes can be
formulated, and, with a little effort, solved.

\bigskip

I have chosen as my thesis topic the investigation of the stellar
normal modes in the plane of a model galaxy. [...
] Some
qualitative features of the equations indicate that the type of
spiral disturbance with two arms is preferred. This result does
not seem to depend critically on the model, which is encouraging.
The final proof has to be left to numerical calculations, which
are not yet complete.'' (Kalnajs 1963, p.1-3) \par}
\end{quotation}

It is seen therefore that Kalnajs was envisaging the disk of stars as a
resonator in which global spiral-wave modes are developed. If \textit{stationary}, the leading
and the trailing components are just mirror-imaged, so that, superimposed,
they give no spiral pattern. However, due to slow non-reversible processes
occurring in real galaxies, the symmetry is violated.

\bigskip

In support of his normal-mode concept, Kalnajs considered large-scale
non-axisymmetric disturbances to a hot inhomogeneous flat stellar disk, and
derived for them a general integral equation whose complicated frequency
dependence implied a discrete wave spectrum. He also pointed out the role of
Lindblad's condition (4). When satisfied, large parts of the galactic disk
could support coherent oscillations for the $m = 2$ mode, whereas for larger
$m$'s there would be Lindblad resonances within the disk. Stars in these
regions feel the perturbing wave potential at their own natural frequency,

\begin{equation}
{\left| {\nu}  \right|} = 1,
\quad
\nu \equiv (\omega - m\Omega ) / \kappa ,
\end{equation}

\bigskip

\noindent thus undergoing strong orbital displacement and making
the $m > 2$ modes lose integrity\footnote{ A combination $\omega -
m\Omega $ is called the Doppler-shifted wave frequency, one
reckoned in a reference system corotating with disk material. The
shift is due to the fact that waves are naturally carried along
by flows.} . Hence Kalnajs concluded that his ``formulation of
the problem'' shows a dynamical preference for two-armed spirals
and ``gives little insight of what to expect in both the shape of
the disturbances and their time dependence when $m > 2$'' (Kalnajs
1963, p.13).

\bigskip

A summarizing exposition of the subject Kalnajs gave in his PhD thesis ``The
Stability of Highly Flattened Galaxies'' presented at Harvard in May 1965
(Kalnajs 1965);\footnote{ Kalnajs' thesis committee members were Layzer, Lin
and Toomre, as officially confirmed from Harvard.} it contained an extended
discussion lavish in ideas and technicalities. At the same time, the thesis
became in fact Kalnajs' official public debut, so that to it as a reference
point should we attach chronology when confronting certain factual points in
the spiral history of the 1960s.

\bigskip
\bigskip

\section*{III. THE LIN-SHU THEORY}
\addcontentsline{toc}{section}{III. The Lin-Shu theory}

\bigskip

{\footnotesize \begin{list}{}{\leftmargin4cm}
\item I would like to acknowledge that Professors Lin and Toomre of MIT are also
interested in the problem of spiral structure, and that I have benefited
from discussions with them as well as their students.
\begin{flushright}
\textit{Kalnajs 1963, p.13}
\end{flushright}
\end{list}}

\bigskip

\subsection*{3.1 Working hypothesis and semi-empirical theory}
\addcontentsline{toc}{subsection}{3.1 \it Working hypothesis and
semi-empirical theory}

\bigskip

{\footnotesize \begin{list}{}{\leftmargin4cm}
\item In hindsight, considering the crucial influence that the Lin {\&} Shu (1964)
paper had on the thinking of astronomers, it is only regretful that Lin did
not decide (with or without me) to publish even earlier, because he
certainly had all the physical ideas contained in our paper well before
1964.
\begin{flushright}
\textit{Shu 2001}
\end{flushright}
\end{list}}

\noindent While Toomre, Hunter and Kalnajs had already presented their first results
in the dynamics of flat galaxies, Lin still kept on thinking over the spiral
problem.\footnote{ Lin's basic themes still were in hydrodynamics (e.g.,
Benney {\&} Lin 1962; Reid {\&} Lin 1963).} Astronomers in Princeton had
convinced him that, despite Chandrasekhar's criticism of Lindblad's
theories,\footnote{ That criticism (Chandrasekhar 1942) concerned only the
asymptotic-spiral theory, and it was itself not flawless as attached to
confusing empirical data of the 1920's -- 30's.} the idea itself of a
long-lived, shape-preserving spiral pattern is consistent with Hubble's
classification system that relates spiral features with a galaxy's
morphological type, its steady characteristic, thus suggesting that the
spirals are steady as well. This view reminded Lin of wave modes in fluid
flows that he had been studying for years back.\footnote{ ``I have been
thinking of modes ever since I learned about the fine points of the Hubble
classification''. (\textit{Lin})} On purely heuristic grounds, discrete spiral modes
seemed to him very reasonable as the natural result of wave evolution, and,  if so, the patterns released might be associated with \textit{slowly growing} or \textit{neutral} modes. Lin raised
this premise to the rank of working hypothesis, and around it as the nucleus
he set to develop a \textit{semi-empirical} theory.\footnote{ ``I adopted the empirical approach
because of my close contacts with the observers (and with Lo Woltjer). Now
that I have thought over the situation some more, I think I should admit
that it is probably true that my past long-standing experience in the
studies of hydrodynamic instability did (as you hinted) play a role in my
thinking (although I was not conscious of it). But more important, I also
feel (upon reflection) that the reason I adopted the empirical approach is
really the natural consequence of my past education. My undergraduate
education was in physics (at Tsinghua University of China, where all the
major professors in Physics had doctorate degrees from English speaking
universities such as Harvard, Caltech, Chicago and Cambridge), with all the
pleasant memories of doing the experiments with precision and the
satisfaction of having the data checked against theory. My graduate
education was primarily at Caltech where I studied under Theodore von
Karman. It is also there that I took a course from Fritz Zwicky who first
identified the regular spiral structure in the Population II objects of the
Whirlpool M51''. (\textit{Lin})} It was seen to follow best the ``urgent assignment from
the astronomers [\ldots ] to make some specific calculations'' and ``to
demonstrate the possibility of the existence of quasi-stationary spiral
modes from the theoretical point of view [\ldots ] with understanding of the
dynamical mechanisms relegated to a secondary and even tertiary position''
(\textit{Lin}).\footnote{ ``Despite of my decades of experience with instability of
shear flows, I did not bring these matters into the presentation of the 1964
paper, but commented only vaguely about instability. [\ldots ] There was no
shortage of theoretical astronomers who understood the mechanisms perhaps
better than I did; e.g. Lo Woltjer and Donald Lynden-Bell and perhaps even
Peter Goldreich (even at that point). Goldreich turned out be the most
successful leader in the understanding of the density waves in the context
of planetary rings''. (\textit{Lin})}$^{,}$\footnote{ ``In hindsight, I think Lin's
judgment was accurate considering how quick people were to attack his point
of view with proofs of `antispiral theorems' and the like shortly after the
publication of LS64''. (\textit{Shu})}

\begin{quotation}{\footnotesize
\noindent ``The conclusion in the working hypothesis is \textit{not proved or deuce}, but supported by an
accumulation of theoretical analysis and empirical data. The adoption of
this working hypothesis is a very important step in the development of a
theory of spiral structure. It means that the authors are committed to back
it up with the comparison of subsequent predictions with observational
data.'' (\textit{Lin}) \par}
\end{quotation}

The coauthor to share Lin's fame and commitment was his student Frank Shu
(Shu 1964)\footnote{ ``All the original ideas were C.C. Lin's, and my
original contributions were mainly to check the equations that he wrote down
and posed as problems. (I did find a way to derive the asymptotic relation
between density and potential by attacking the Poisson integral directly,
but even there I initially blundered in not realizing the necessity of an
absolute value on the radial wavenumber. The final derivation presented in
the appendix of LS64 is due to Lin). I did considerable reading, however, on
the astronomical side and may have contributed some ideas concerning how OB
stars form and die in spiral arms. (This was the beginning of my lifelong
interest in star formation.) Lin was indeed quite generous to include me as
a coauthor on LS64, and I will always be grateful for his guidance and
support of a young (I was 19 at the time) undergraduate student''. (\textit{Shu})} who
``found it remarkable that a scientist trained as a professional
mathematician would place higher priority on empirical facts than deductive
reasoning'' and believed that ``it was this broad-mindedness and clear
vision that gave Lin a considerable advantage over his many competitors of
the period'' (\textit{Shu}).\footnote{ ``Lin undoubtedly encouraged many of his younger
colleagues -- like Alar Toomre -- to think about the problem of spiral
structure. I can only imagine that Lin's treatment of people then much more
junior than himself was equally as generous as his treatment of myself.
Certainly, he must have discussed with Alar Toomre (and later Chris Hunter)
his ideas about this problem. Toomre's early papers on the subject
acknowledge this debt of introduction and inspiration. Why then did those
early papers not carry Lin's name as a coauthor? I do not know, nor would I
dare to probe (by asking either Lin or Toomre) for fear of opening old
wounds that are best left closed''. (\textit{Shu})\par One way or another, no alliance
was formed between Lin and Toomre. They ``diverged in emphasis from the very
beginning'', so that ``there were discussions, but no real collaboration''
(\textit{Lin}). As in agreement with this Toomre recalls that back again at MIT in
spring 1963 he did decline Lin's ``astonishing suggestion to write some such
paper jointly, since he himself had contributed almost nothing very
concretely to my gravitational (in)stability insights, and yet also since I
likewise felt I had added next to nothing to his own spiral-wave hopes''
(\textit{Toomre}).} The Lin and Shu paper ``On the spiral structure of disk galaxies'' (Lin
{\&} Shu 1964, hereinafter LS64), in which ``they first demonstrated the
plausibility of a purely gravitational theory for density waves by a
continuum treatment'' (Lin {\&} Shu 1966, p.459), appeared in August
1964.\footnote{ That the historical Lin {\&} Shu article was referred to as
`Lin's (1963) preprint' by Layzer (1964) and as `Lin (1964)' by Toomre
(1964) and Kalnajs (1965) as it was about to appear in the fall of 1964
speaks of its urgently extended coauthorship as Lin's last moment decision
(so striking for a well-motivated and ambitious scientist).\par Anyway, the
Lindblad (1964) paper, also considering quasi-stationary circulation and the
resulting spirals in differentially rotating galaxies, appeared half a year
\textit{prior} to Lin's patent. The authors had neither contacts nor fresh news on each
other's most parallel work, and hardly could have it. ``There was no
justification to trouble B. Lindblad with a novice being converted, Lin
explains. I was waiting for a definitive new prediction before writing to
him. Even then I would have done it through P.O. Lindblad for several
obvious reasons. Unfortunately, by the time our result came out (IAU
Symposium No 31) [see Sect.3.2] he already passed away'' (\textit{Lin}). Even less
probable was any contact-making step from the other side. ``About that time
[fall of 1964] my father was on a trip around the world caused by the
inauguration of the Parkes telescope in Australia, P.O. Lindblad recalls. On
his way home he passed through the US [\ldots ] but he brought no news about
density wave theories. [\ldots ] I think my father was aware of the
existence of the LS64 paper but had not had the time to penetrate it. I know
that he was happy to learn from Whitney Shane, who visited us around the
beginning of June 1965, that his work on spiral structure had been more and
more appreciated recently''. (\textit{P.O. Lindblad})} $^{} $

\bigskip

The paper considered small non-axisymmetric disturbances to a razor-thin
cold disk and found for them, through the governing hydrodynamic and Poisson
equations, wave-like solutions of the type

\begin{equation}
\psi (r,\theta ,t) = Re{\left\{ {\varphi (r)\exp [i(\omega t - m\theta ]}
\right\}},
\quad
\varphi (r) \equiv A(r)\exp {\left[ {iS(r)} \right]},
\end{equation}

\bigskip

\noindent each specified by its eigenfunction $\varphi (r)$ and a
pair of eigenvalues $\omega $ and $m$. For further advancement,
the WKBJ-method was applied. It is valid for the case of phase
$S(r)$ varying with radius much faster than amplitude $A(r)$,
which features the \textit{tightly wrapped} spirals, ones of small
pitch angle between the circumferential tangent and the tangent
to the constant-phase line

\begin{equation}
\omega t - m\theta = const.
\end{equation}

\bigskip

Depending on the sign of a radial-wavenumber function $k(r) = - \partial S /
\partial r$, the spirals are trailing ($k > 0)$ or leading ($k < 0)$ (Fig.8).
With $A(r)$ expanded in a series over a small parameter tan$i = m
/ kr$ ($i $ being the pitch angle), the problem is solved to the
lowest, $i$-independent order neglecting the azimuthal force
component of spiral gravity. In this case, both leading and
trailing arms act as just rings, so that the ensuing dispersion
relation

\begin{equation}
\nu ^{2} =
1 - {\left| {k} \right|} / k_{T} ,
\quad
\nu \equiv (\omega - m\Omega ) / \kappa
\end{equation}

\bigskip

\noindent substantially repeats Toomre's equation (5) for radial
oscillations. Importantly, relation (14) is valid for Re{\{}$\nu
^{2}${\}}$ \le $1. This restricts the radial span of the WKBJ
solutions, and in the neutral case Im{\{}$\nu ${\}}$ = 0$ they
gain the territory between the Lindblad resonances determined by
Eqn (11) and equating the angular speed of an $m-$armed spiral
pattern to a combination

\begin{equation}
Re\{\omega / m\} \equiv \Omega _{p} = \Omega (r) \mp
{\frac{{\kappa (r)}}{{m}}}
\end{equation}

\bigskip

\noindent with the minus/plus sign discriminating, respectively,
between the ILR and OLR. The two-armed spirals thus seem
preferred as best covering an entire disk (Fig.9).

\begin{figure}
\centerline{\epsfxsize=0.8\columnwidth\epsfbox{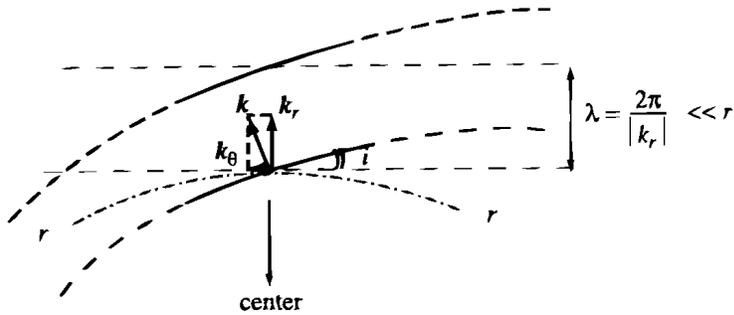}}
\caption{\footnotesize \textit{The WKBJ approximation and the
tightly wrapped spiral waves}. $k_{r} \equiv k$ and $k_{\theta}
\ll k$ are the components of the local wavenumber
\textbf{\textit{k}}. $\lambda = 2\pi / k$ determines the radial
interarm spacing; it is small compared to the galactocentric
distance $r$ since $kr \gg 1$ (which is equivalent to small pitch
angles $i \ll 1$).}
\end{figure}

\bigskip

Such was the mathematical basis of the original Lin-Shu density-wave theory,
called elementary by its authors any later (e.g. Bertin {\&} Lin 1996, p.229). It treated wave quantities $\Omega _{p} $, $\gamma $, and $m$ as free
parameters burdened with no dynamical imposition, which made the theory so
comfortable in imitating spiral grand designs by means of the curves
$r(\theta )$ given by

\begin{equation}
m(\theta - \theta _{0} ) = - {\int\limits_{r_{0}} ^{r} {k_{T} Re\{1 - \nu
^{2}\}dr}}
\end{equation}

\bigskip

\noindent and obtained through the integration of expressions (13)
and (14). Sure, the results of this procedure were controvertible,
already because the \textit{fast}-growing waves -- exactly those
examined in LS64 -- ruled out the proclaimed
quasi-stationarity.\footnote{ To soundly fit the empirical 2-3
kpc local-arm spacing in the Milky Way, LS64 chose a combination
of angular speed $\Omega _{p} = 10$km/s/kpc and growth rate
$\gamma = 50$km/s/kpc (!) for their tentative two-armed spiral.}
But the authors hoped that random motions, excluded from their
analysis, would in fact stave off disk instability as
definitively as to impose a state of near-stability open for
\textit{slowly} growing modes until a small but finite amplitude.

\begin{figure}[t]
\centerline{\epsfxsize=0.7\columnwidth\epsfbox{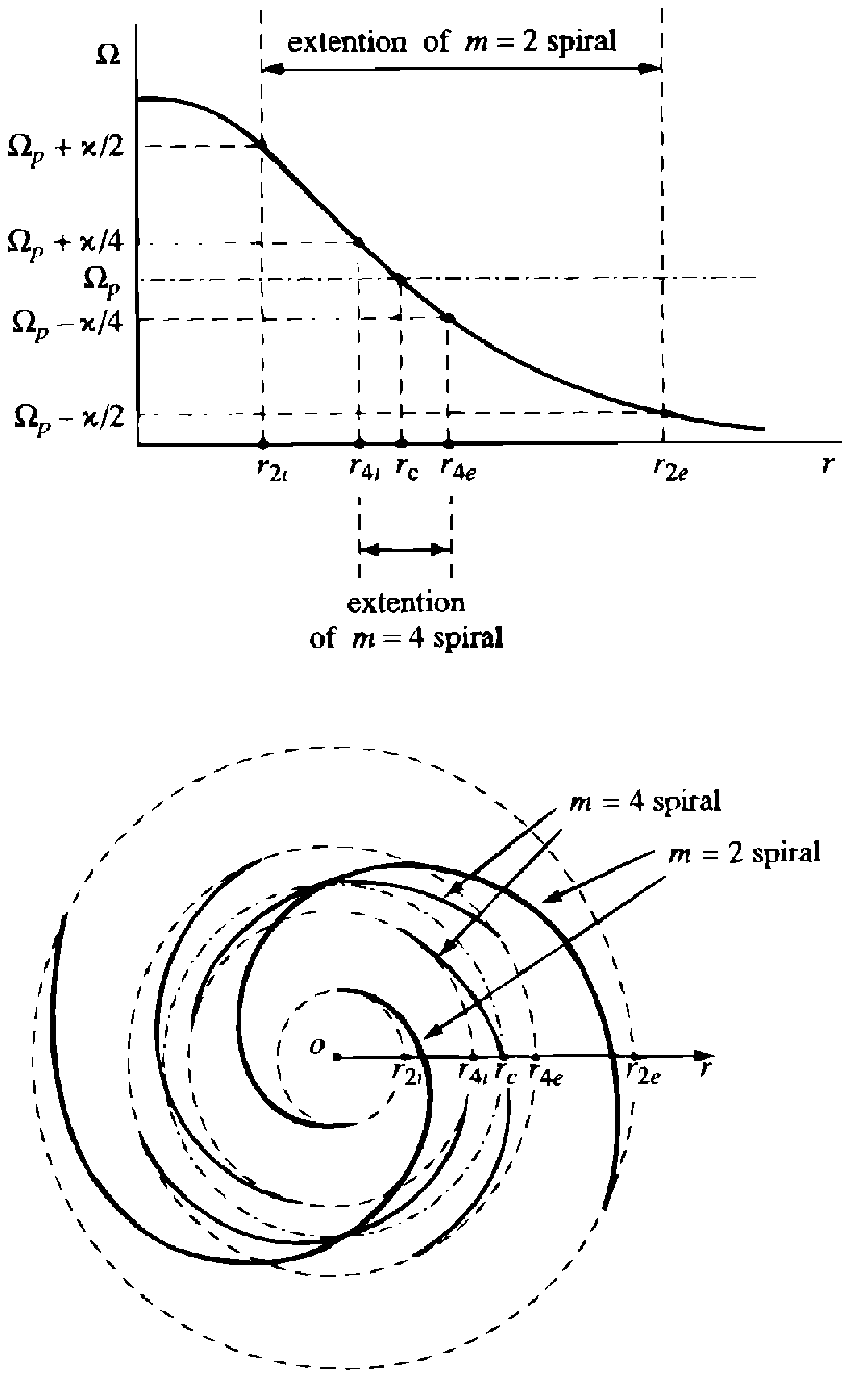}}
\caption{\footnotesize \textit{The Lindblad resonances as
confining the region accessible for the tightly wrapped spiral
waves}. ($a$) -- a rotation curve for a galaxy disk and its
corresponding corotation and $m$ = 2, 4 Lindblad resonances;
($b$) -- the co-scaled view of the two and four armed tightly
wrapped spirals.}
\end{figure}

\bigskip
Toomre (1964) had reflected already on such a state of $Q \cong 1$ as
settling \textit{once} all over the disk-like stellar Galaxy, but yet he found it stable\textit{ still},
at least in our solar region. As a counterpoise, Lin with Shu diagnosed
instability for another region, at about $r_{0} = 4-5$ kpc from the
center. With that, they pictured ``a galactic disk, which is in part stable
and in part unstable'' and suggested ``the possibility of a balance
resulting in a neutral density wave extending over the \textit{whole} disk and having a
scale of the order of (but smaller than) the distance between the stable and
unstable regions'' (LS64, p. 651). It was this ``suggestion of the
possibility'' that summarized Lin's early reflections and made his basic
working hypothesis originally sound as a statement that

\begin{quotation}{\footnotesize
\noindent ``the total stellar population, which has various degrees of velocity
dispersion, forms a \textit{quasi-stationary spiral structure} in space of the
general nature discussed above'' (LS64, p.651). \par}
\end{quotation}

\bigskip 

As we can see, this statement hinges almost entirely on the opinion that,
for our galactic disk to be equally stable at that $r_{0} $, the velocity
dispersion must there exceed $c_{r,\min}  \cong 80\pm 10$ km/s, which
cannot be the case, else ``a considerable number of stars with high radial
velocities would reach our neighborhood from the interior part of the
Galaxy, contrary to observational evidence'' (LS64, p.651). But was this
opinion (the authors never repeated it) strong enough? First, it meant an
inconceivable situation when some \textit{massive} portion of a stellar galaxy remains
\textit{unstable} during all the period of formation in it of a global quasi-\textit{steady} pattern.
Secondly, and most important for astronomers, it had -- already in 1964 --
grave objections to the fact that the largest epicyclic deflection of the
Lin-Shu ``high radial velocity stars'' from their `home' radius $r_{0}  =
4-5$ kpc, equaled to $\Delta r \cong r_{0} c_{r} / V_{0} \sqrt {2} $, was
in frames of Schmidt's model (cited in LS64) $1-1.5$ kpc only -- too
little to let those stars even come close from $r_{0} $, if not reach us. We
find that the original QSSS hypothesis of Lin and Shu, called nowadays ``a
preliminary formulation'' \textit{only} (Bertin {\&} Lin 1996, p.80), rested on a rather weak basis, both dynamical and empirical.$^{} $

\bigskip

Very interesting in LS64 is the authors' notice on what had made their work
get to print so urgently. A passage following their opening discussion of
``at least two possible types of spiral theories'', one of which ``is to
associate every spiral arm with a \textit{given body of matter}'' and the other ``is to regard the spiral
structure as a [quasi-steady] \textit{wave pattern}'', reads:

\begin{quotation}{\footnotesize
\noindent ``Toomre tends to favor the first of the possibilities described above. In
his point of view, the material clumping is periodically destroyed by
differential rotation and regenerated by gravitational instability.\footnote{
``The prevalent thinking among the other prominent theorists of the time --
and this included Alar Toomre -- was that spiral structure was a chaotic and
regenerative phenomenon -- `shearing bits and pieces', as Alar later put it
in one of his papers''. (\textit{Shu})} [\ldots] The present authors favor the second point of view [\ldots] Since A. Toomre's (1964) point of view has been
published, it seems desirable to publish our point of view even though the
work is not yet as complete as the present writers would wish to have it.''
(LS64, p.646) \par}
\end{quotation}

\noindent This puzzles. Although it is true that from about 1962 onward Toomre
suspected -- much as Lynden-Bell had already done in his thesis two years
earlier, as it turned out -- that at least the more ragged-looking spiral
structures result primarily from recurrent gravitational instabilities in
the plainly dissipative gas layer of a galaxy (\textit{Toomre}), there was no explicit
discussion of any such suspicions in T64 as actually published. One cannot
help but think that this accentuated mention of `Toomre (1964)' was more
than just a mistaken reference, that actually it betrayed the influence that
at least the cited paper had on Lin.

\begin{quotation}{\footnotesize
\noindent \textit{Shu}: ``Here, I can only speculate, because certainly my foresight then was not
as sharply developed as Lin's. Nor was I privy to the developing
estrangement between him and Alar Toomre. [\ldots ] Lin had been thinking
about the problem of spiral structure nonstop since the Princeton conference
in 1961. But he had a world-renowned reputation to protect and therefore was
loathe to publish anything hasty before he had worked out his ideas
mathematically to his satisfaction. [\ldots ] Lin (and later, I) felt
strongly that spiral structure was, in essence, a normal mode. But by all
the standards of what was then known, a normal mode could not be spiral
(unless it grew ridiculously fast). Nevertheless, Lin felt sure that one
should not do the naive thing of superimposing equal trailing and leading
parts when the wave frequency is (nearly) real. And he probably wanted to
discover the reason why before publishing anything. Alar's 1964 paper
triggered him into premature action''. (\textit{Shu})

\bigskip

\noindent \textit{Lin}: ``The urgency in my submittal of our paper was to present a \textit{different} perspective,
not to fight for priority''. ``After reviewing the paper again, I think I
could not have done much better or even any better''. (\textit{Lin}) \par}
\end{quotation}

\bigskip

One way or another, we see that by 1964 Lin indeed had had several thoughts
and feelings about spiral modes, and he was eager about gaining power to his
perspective. At that, he knew of a growing optimism with shearing or
evolving density waves\footnote{ Goldreich and Lynden-Bell in England and
Julian and Toomre at MIT set to work on this by 1964.} and, as well, of the
parallel wave-mode interest at Harvard. The T64 paper\footnote{ The revised
version of T64 was submitted in January 1964.} , apart from its engagements
on disk stability, did mention Kalnajs' advancing efforts and, still more
glaringly, it also mentioned and already \textit{discussed} Lin's yet unpublished solutions.\footnote{
Toomre concluded that ``whatever differences there may exist between the
shorter axisymmetric and non-axisymmetric disturbances, these must in
essence be due only to the circumstance of \textit{differential} rotation'' (T64, p.1223). In
Lin's hands, in contrast, this `circumstance' still allowed the dispersion
relation (14) for non-axisymmetric waves to be rather close to its
axisymmetric analog (5), although the waves stood as steady-mode solutions
of the WKBJ type. Yet, as well, the governing equations admitted an
``altogether different family of approximate non-axisymmetric solutions''
(T64, p.1223), with the radial wavenumber proportional to the disk shear
rate $A $(Oort's constant), and growing with time, $k_{r} \propto At$. This
meant that a spiral disturbance of the leading form ($t < 0)$ unwrapped,
started trailing, and then wrapped tighter and tighter ($t > 0)$. Thus the
point was that, on the one hand, differential rotation continuously deforms
even the tightly-wound spiral waves of this sort, whereas, on the other
hand, these ``should probably be regarded as particular superpositions of
Lin's solutions'' (T64, p.1223). This discordance was thought to be removed
by a fuller analysis beyond the WKBJ-limit.}  This must have put Lin in a
position to urgently patent his views, albeit makeshift in argument for want
of better mathematics, and in so doing he rather awkwardly exhibited the
opponents' preoccupations as an alternative already placed on record.

\vfill\break

\subsection*{3.2 A definitive (?) new prediction}
\addcontentsline{toc}{subsection}{3.2 \it A definitive (?) new prediction}

\bigskip

{\footnotesize \begin{list}{}{\leftmargin4cm}
\item A desirable feature of the WKBJ waves is their mathematical simplicity;
their physical relevance to the `grand design' of a spiral galaxy is less
transparent.

\begin{flushright}
\textit{Kalnajs 1971, p.275}
\end{flushright}
\end{list}}

\bigskip

\noindent ``Just how much did Kalnajs' study of axisymmetric
oscillations influence our work? The simple answer is: very
little, if at all'' (\textit{Lin}). Such is Lin's judgment
regarding the results he had set out in the summer of
1965.\footnote{ Lin presented his first hot-disk results in June
1965 at a summer school at the Cornell University and at a
mathematical symposium at the Courant Institute. These materials
were published in two extensive articles (Lin 1966, 1967a)
submitted in July. ``I recall becoming aware of the relationship
with the work of Kalnajs only when he brought up the issue in
connection with Frank Shu's thesis presentation. I immediately
recognized that there would probably be a way to make the
connection through the application of the Mittag-Leffler theorem.
Note that it is easy to derive the Kalnajs form from our integral
form, but difficult to reverse the process. And our numerical
calculations depended on the simple integral, since it was a time
when large scale use of the computer was not yet available in a
mathematics department. (I still remember the painful experience
when my request -- as chairman of the committee on applied
mathematics -- for a computer was turned down, even though the
department had the funds. [\ldots ] Kalnajs might have been able
to check the calculations with his infinite series through the
use of the computer.)'' (\textit{Lin})} Those got out of the
printer in no less than one year (Lin 1966, 1967a), but an
abridged and slightly updated version appeared as soon as
February 1966, having become an ``Outline of a theory of density
waves'' by Lin and Shu (1966), labeled `Paper II'.

\bigskip

The three issues reported a WKBJ-styled dispersion relation for the
razor-thin hot disk,

\begin{equation*}
\nu ^{2} = 1 - {{{\left| {k} \right|}} \mathord{\left/ {\vphantom {{{\left|
{k} \right|}} {k_{T} \cdot F_{\nu}  (x),}}} \right.
\kern-\nulldelimiterspace} {k_{T} \cdot F_{\nu}  (x),}}
\end{equation*}

\begin{equation}
F_{\nu}  (x) = {\frac{{1 - \nu ^{2}}}{{x}}}{\left[ {1 - {\frac{{\pi \nu
}}{{\sin \pi \nu}} }{\frac{{1}}{{2\pi}} }{\int\limits_{ - \pi} ^{\pi}  {e^{
- x(1 + \cos s)}\cos \nu sds}}}  \right]}.
\end{equation}

\bigskip

\noindent From its Kalnajs' axisymmetric analog (9)-(10) it differed in the Doppler
shift included in $\nu $ and in the form of the reduction factor $F_{\nu}
(x)$.\footnote{ ``I have little knowledge but I make this conjecture:
Kalnajs was studying axisymmetric oscillations, not standing waves of the
spiral form, and obtained his results through the use of results for
analogous oscillations in plasma waves. (I learned a lot about plasma
physics only after Y.Y. Lau joined our research group.)'' (\textit{Lin})} It was an idea
of some such dispersion relation, Lin and Shu (1966) remarked, that had fed
originally (LS64) their insight in the disk-stabilizing role of random
motions.\footnote{ Lin agreed that the dispersion relation was already
derived by Kalnajs ``in the special case of axially symmetrical
disturbances'', but ``by a quite different method'' and ``independently of
the work of the author'' (Lin 1966, p.902). He certainly appeared rather
sensitive on the point of independence, beginning his spiral studies. His
\textit{first} appraisal of Lindblad's long-term emphasis on steady spirals was: ``Indeed,
independently of each other, B. Lindblad (1963) and the present writer came
to the same suggestion of a \textit{quasi-stationary spiral structure of the stars} in a disk galaxy'' (Lin 1966, p.898). Again,
referring time and again to different methods adopted by him and his various
competitors, Lin found it difficult to closely compare those related issues.
But, for example, Lynden-Bell (1962) and Toomre (1964) had used the same
characteristics method as that taken in 1965 by Lin, with which he basically
re-derived, again independently, this time from Toomre, that crucial
differential equation of `asymptotic' disk-stability and density-wave
theories (cf. Eqn (53) from T64 with Eqn (7.15) in Lin 1966 and Eqn (ю20) in
Lin et al 1969), not having mentioned its factual use by his next-door
institute colleague.}

\bigskip

But an important dynamical, not chronological, point was that the hot
rotating disk was seen to conduct radial and spiral waves rather distinctly.
Given a state of marginal stability, the oscillatory radial neutral mode
$\nu = \omega / \kappa = 0$ is well maintained by it along its medium radii
(dying out at large $r$'s),\footnote{ Such behavior is well seen on Fig.3 from
T64 showing results of numerical calculations of global radial modes for
some illustrative cold-disk model.} the local wavelength function $\lambda
_{0} (r)$ depending on mass and angular momentum distributions. In contrast,
the spiral wave cannot be neutral as extendibly: its Doppler-shifted
frequency $\omega - m\Omega (r)$ gets $r-$dependent. This ties the neutrality
condition $\nu = (\omega - m\Omega ) / \kappa = 0$ to a narrow corotation
zone of $r \cong r_{c} $, and there only can the interarm spacing $\lambda
(r)$ equal $\lambda _{0} (r)$, the rest of disk getting more and more stable
against the wave as one travels away from $r_{c} $ in or out. If so, why not
to try to juxtapose the basic Lin-Shu concept of a balance and the
solar-region stability inference by Toomre? For this, it seems sufficient to
send corotation way beyond -- to an outer disk region supposedly as
permissive to marginal stability as to admit it -- and to cancel all
instability inside that $r_{c} $ in favor of $Q \ge 1$. Lin and Shu did seem
to have followed this way. Moreover, they adopted a $Q \equiv 1$ model
(discussed already in T64), being captured by a picture of overstability,
i.e. gradient instability held to mildly develop over the system and to
provide some selective amplification of trailing, not leading, waves.

\begin{figure}[t]
\centerline{\epsfxsize=0.5\columnwidth\epsfbox{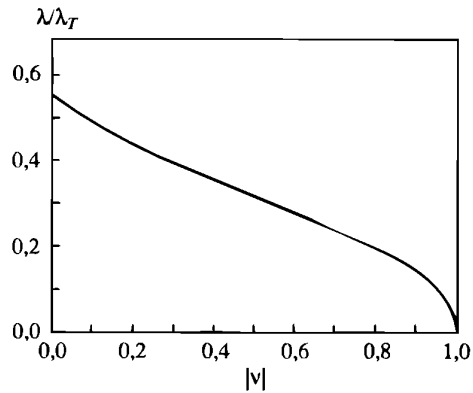}}
\caption{\footnotesize \textit{The short-wave branch of the dispersion
relation (17) for a Q = 1 disk model}. (The figure is
reproduced from Lin {\&} Shu 1967)}
\end{figure}

\bigskip

Besides, relation (17) tells $\nu (k)$ to decrease with
wavenumber till $k$ remains under some $k_{0} $, and then to rise
up at $k \to \infty $ back to unity. Any intermediate value of
$\nu $ is met thus twice, meaning two branches of WKBJ solutions,
the shorter- and the longer-wave ones, their forms $r(\theta )$
being provided by equation (16) with $F_{\nu}  (x)$ added in the
integrand denominator. If $Q \equiv 1$, the branches join at
corotation, showing there equal interarm spacings $\lambda _{sw}
(r_{c} ) = \lambda _{lw} (r_{c} ) = \lambda _{0} (r_{c} )$. This
value is the largest (smallest) for the shortwave (longwave)
branch: $\lambda _{sw} (r)$ falls down until zero ($\lambda _{lw}
(r) \to \infty )$ as one goes from corotation to ILR. Aimed from
the outset at explaining the observed 2-3 kpc local spacings, Lin
got tempted to acknowledge the shortwave branch, the more so as,
not to forget, in 1964 he had had no choice when having to
comment on this same gas-given spacing on the basis of relation
(14) that seized but \textit{one} -- long-wave (!) --
branch.\footnote{ LS64 had assumed that because not all the stars
but only those with smallest random velocities perceptibly
contribute to the response of a disk, its effective surface
density must be several times less than its full value.} But
things did not get all as clear by 1966, and this is why neither
Lin (1966, 1967a) nor Lin and Shu (1966) were eager to go into
the wave-branch question, keeping silent about any graphic view
of their newer formula. Only at the Noordwijk IAU Symposium
(August, 1966) they gave a graph, it displayed the
\textit{short}-wave-branch extension of the $\lambda (\nu )$
curve (Fig.10) on which they built a model for the full spiral of
our Galaxy (Fig.11), tentatively two-armed and answered by a
remote corotation (Lin {\&} Shu (1967).\footnote{ ``This was my
first meeting with the distinguished astronomers who made all the
important observations related to spiral structure, many of whom
worked under Oort's direction. Here we presented our first
prediction of the spiral structure of the Milky Way, which
remained to be an approximate representation, as indicated by
Yuan's continual refinement over the years''. (\textit{Lin})}
Spirals of this class show as slow a rotation as to almost
guarantee the ILRs be present and lie in a relative proximity
from the center. Namely, Lin and Shu connected our `home' $m $= 2
ILR with the `3-kpc arm' which fixed the spiral pattern speed $\Omega _{p}
= 11$ km/s/kpc.

\begin{figure}[t]
\centerline{\epsfxsize=0.5\columnwidth\epsfbox{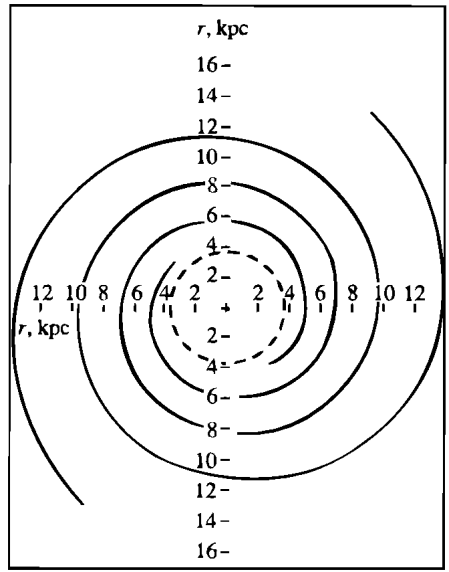}}
\caption{\footnotesize \textit{The Lin-Shu model for the Galactic
spiral density wave}. The model is calculated with the help of
the dispersion curve in Fig.10. The dashed line shows the ILR
region taken to be the residence of the `3-kpc arm'. This
provides the pattern speed $\Omega _{p} = 11$ km/s/kpc. (The
figure is reproduced from Lin {\&} Shu 1967)}
\end{figure}

\begin{quotation}{\footnotesize
\noindent ``My earliest recollection of realizing that there were
separate long and short branches came when I was doing the
numerical calculations for the spiral pattern that Lin wished to
show at the Noordwijk symposium. As I recall, he was in the
Netherlands and I remained behind at Harvard, and we corresponded
by mail. I was considerably confused by which of the two branches
should be used to generate spiral patterns (I had realized that a
`reduction factor' applied to our 1964 formula was an incomplete
description, and that long and short waves were implicit to
Toomre's evaluation of a critical $Q$ for axisymmetric
disturbances). Finally, Lin suggested that we should simply
choose the short branch by fiat as the practical thing to do
given the press of the Noordwijk presentation, and we were left
to try to sort things out later. That's my memory of the
events''.\footnote{ ``Lin and Shu 1966 emphasis upon (and the
dispersion relation for) the \textit{short-wave} branch of nearly
axisymmetric WKBJ-style density waves, which is something that
Kalnajs (1965) also knew from his thesis but failed to emphasize
nearly as adequately, escaped me altogether even though the same
for the \textit{long-wave} branch as well as the stability
criterion were plain as day from T64 -- and to a more limited
extent even from Safronov (1960a,b), as I often agreed in
retrospect. I think my trouble was that my own ongoing work then
with Julian (Julian {\&} Toomre 1966) [...] had also sensitized me
to the severity of \textit{phase mixing}. [...] Looking back, this
made me suspect until well into 1965 that all \textit{short}
stellar-dynamical waves, unlike their over-idealized gas
equivalents, would in fact be strongly damped and were probably
not of much value. And right there I have cheerfully agreed for
about 34 years now that Lin and Shu (and as an independent
authority also Kalnajs, not at all to be omitted) together proved
me to have been spectacularly \textit{wrong}''.
(\textit{Toomre})} (\textit{Shu}) \par}
\end{quotation}

\nobreak

The Noordwijk diagram has been the first presentation of our Milky Way's
density wave.

\goodbreak

\bigskip

\section*{Afterword}
\addcontentsline{toc}{section}{Afterword}

\bigskip

As we have seen here, understanding the spiral structure of galaxies took
many twists and turns even in the hands of Bertil Lindblad who seems rightly
regarded the main father of this whole subject. By the early 1960s, with the
arrival of computers, plasma physics and several fresh investigators, it
entered a new period of unusually vigorous activity, not always very united
or monothematic, but broadly grouped under the umbrella marked `density-wave
theory'. Its foremost enthusiast and proponent was undoubtedly C.C. Lin,
whose 1964 and 1966 papers with Shu had a big and immediate impact upon
other astronomers, at least as a welcome sign that genuine understanding of
the spiral phenomenon seemed in some sense to be just around the corner.

\bigskip

In retrospect, even Lin occasionally let himself get carried away with too
much enthusiasm as for instance when he wrote in his 1967 review article
that his relatively exploratory work with Shu had already led to a ``theory
free from the kinematical difficulty of differential rotation'', or that it
``enables us to provide a mechanism to explain the existence of a spiral
pattern over the whole disk while allowing the individual spiral arms to be
broken and fragmentary'' (Lin 1967b, p.462). Already at the time such
optimism was not entirely shared by other experts. And by the late 1960s --
as we shall see in Paper II -- it had become very clear to everyone that
much hard work still remained to explain even the persistence, much less the
dynamical origins, of the variety of spirals that we observe.

\newpage

\section*{References}
\addcontentsline{toc}{section}{References}

\bigskip

{\footnotesize

(\textit{Chandrasekhar}) = S. Chandrasekhar. Private communication, 1982.
\noindent (\textit{Contopoulos}) = G. Contopoulos. Private communications, 2000-01.

\noindent (\textit{Hunter}) = C. Hunter. Private communications, 2000-01.

\noindent (\textit{Kalnajs}) = A. Kalnajs$.$ Private communications, 2001.

\noindent (\textit{P.O. Lindblad}) = P.O. Lindblad. Private communications, 2000-01.

\noindent (\textit{Lebovitz}) = N. Lebovitz. Private communications, 2001.

\noindent (\textit{Lin}) = C.C. Lin. Private communications, 1982, 2000-01.

\noindent (\textit{Toomre}) = A. Toomre. Private communications, 1982, 2000-01.

\bigskip

\noindent ApJ = Astrophysical Journal

\noindent Ark = Ark. Mat. Astron. Fys.

\noindent MN = Monthly Notices of the Royal Astron. Soc.

\noindent SOA = Stockholm. Obs. Ann.

\bigskip
\bigskip
{\leftskip=0.25in\parindent=-0.25in

Ambartsumian, V.A. 1938. \textit{On the dynamics of open clusters}. Uchenye Zap. Leningr. Gos. Universiteta No 22,
math. sci. series (astronomy), \textbf{4}, 19-22.

Antonov, V.A. 1960. \textit{Remarks on the problem of stability in stellar dynamics}. Astron. Zh. \textbf{37}, 918-926.

Baade, W. 1963. \textit{Evolution of stars and galaxies}. Cambridge MA, Harvard Univ. Press.

Babcock, H.W. 1939. \textit{The rotation of the Andromeda nebula}. Lick Obs. Bull. \textbf{19}, No 498, 41-51.

Bel, N., Schatzman, E. 1958. \textit{On the gravitational instability of a medium in non-uniform rotation}. Rev. Mod. Phys. \textbf{30}, 1015-1016.

Benney, D.J., Lin, C.C. 1962. \textit{On the stability of shear flows}. Proc. Symp. Appl. Math. \textbf{13}:
Hydrodynamic Instability.

Bernstein, I.B. 1958. \textit{Waves in plasma in a magnetic field}. Phys. Rev. \textbf{109}, 10.

Bertin, G., Lin, C.C. 1996. \textit{Spiral structure in galaxies. A density wave theory}. Cambridge MA, MIT Press.

Binney, J., Tremaine, S. 1987. \textit{Galactic dynamics}. Princeton NJ, Princeton Univ. Press.

Bok, B.J., Bok, P.F. 1957. \textit{The Milky Way}. 3$^{rd}$ ed. Cambridge MA, Harvard Univ. Press.

Brown, E.W. 1925.\textit{ Gravitational forces in spiral nebulae}. ApJ \textbf{61}, 97-113.

Burbidge, E.M. 1971. \textit{The evolution of spiral structure.} In: Structure and evolution of the Galaxy. L.N.
Mavridis. Ed. D. Reidel Publ. Comp., 262-283.

Burbidge, E.M., Burbidge, G.R., Prendergast, K.H. 1959. The rotation and mass
of NGC 2146. ApJ \textbf{130}, 739-748.

Burbidge, G.R. 1962. \textit{Evolution of galaxies}. In: The distribution and motion of interstellar
matter in galaxies (Proc. Conf. Inst. Adv. Study, Princeton, 1961). L.
Woltjer, ed. W.A. Benjamin, NY, 291-303.

Chandrasekhar, S. 1942. \textit{Principles of stellar dynamics}. Univ. of Chicago Press, Chicago, IL.

Chandrasekhar, S. 1943. \textit{Stochastic problems in physics and astronomy}. Rev. Mod. Phys. \textbf{15}, No 1, 1-89.

Chandrasekhar, S. 1953. \textit{Problems of stability in hydrodynamics and hydromagnetics} (G. Darwin Lecture). MN \textbf{113}, 667-678.

Chandrasekhar, S. 1969. \textit{Ellipsoidal figures of equilibrium}. New Heaven, Yale Univ. Press.

Contopoulos, G. 1958. \textit{On the vertical motions of stars in a galaxy}. SOA \textbf{20}, No 5.

Contopoulos, G. 1960. \textit{A third integral of a motion in a galaxy}. Zeitschr. f. Astrophys. \textbf{49}, 273-291.

Contopoulos, G. 1972. \textit{The dynamics of spiral structure. Lecture notes}. Astronomy Program, Univ. of Maryland.

Dekker, E. 1975. \textit{Spiral structure and the dynamics of flat stellar systems}. PhD thesis, Leiden Univ.

Efremov, Yu.N. 1989. \textit{Sites of star formation in galaxies}. Moscow, Nauka Publ. [\textit{in Russian}].

Einstein, A. 1953. \textit{Foreword} in: Galileo Galilei. Dialogue concerning the two chief
world systems, Ptolemaic and Copernican. Berkeley.

Fricke, W. 1954. \textit{On the gravitational stability in a rotating isothermal medium}. ApJ \textbf{120}, 356-359.

Genkin, I.L., Pasha, I.I. 1982. \textit{On the history of the wave theory of spiral structure}. Astron. Zh. \textbf{59}, 183-185.

Gingerich, O. 1985. \textit{The discovery of the spiral arms of the Milky Way}. In: The Milky Way Galaxy (Proc. IAU Symp. No 106), H.
van Woerden et al. (eds)., 59-70.

Hubble, E. 1936. T\textit{he realm of the nebulae}. New Haven, Yale Univ. Press.

Hubble, E. 1943. \textit{The direction of rotation in spiral nebulae}. ApJ \textbf{97}, 112-118.

Hulst, H.C. van de, Muller, C.A., Oort, J.H. 1954. \textit{The spiral structure of the outer part of the galactic system derived from the hydrogen emission at 21 cm wave length}. Bull. Astron. Inst.
Netherl. \textbf{12}, No 452, 117-149.

Hulst H.C. van de, Raimond, E., Woerden, H. van 1957. \textit{Rotation and density distribution of the Andromeda nebula derived from observations of the 21-cm line}. Bull. Astron. Inst.
Netherl. \textbf{14}, No 480, 1-16.

Hunter, C. 1963. \textit{The structure and stability of self-gravitating disks}. MN \textbf{126}, 299-315.

Hunter, C. 1965. \textit{Oscillations of self-gravitating disks}. MN \textbf{129}, 321-343.

Idlis, G.M$. $1957. \textit{The cosmic force fields and some problems of structure and evolution of the galactic matter}. Izv. Astrofiz. Inst. AN Kazakh. SSR. \textbf{4}, No 5-6,
3.

Jeans, J.H$. $1929. \textit{Astronomy and cosmogony}. 2$^{nd}$ ed. Cambridge Univ. Press.

Julian, W.H., Toomre, A. 1966. \textit{Non-axisymmetric responses of differentially rotating disks of stars}. ApJ \textbf{146}, 810-830.

Kalnajs, A.J. 1962. \textit{Stellar kinematics}. A research examination submitted by Agris J. Kalnajs
(Jan 31, 1962).

Kalnajs, A.J. 1963. \textit{Spiral Structure in Galaxies}. \textit{Outline of a thesis}. Harvard Univ., Cambridge, MA (Oct 25, 1963).

Kalnajs, A.J. 1965. \textit{The Stability of Highly Flattened Galaxies}. PhD Thesis. Dept of Astron., Harvard Univ., Cambridge,
MA (April 1965).

Kalnajs, A.J. 1971. \textit{Dynamics of flat galaxies. I}. ApJ \textbf{166}, 275-293.

Kurth, R. 1957. \textit{Introduction to the mechanics of stellar systems}. London - NY - Paris, Pergamon Press.

Kuzmin, G.G. 1952. \textit{On the mass distribution in the Galaxy}. Publ. Tartu Obs. \textbf{32}, No 4, 211.

Kuzmin, G.G. 1956. \textit{A stationary Galaxy model admitting triaxial velocity distribution}. Astron. Zh. \textbf{33}, 27-45.

Kwee, K.K., Muller, C.A., Westerhout, G. 1954. \textit{The rotation of the inner parts of the Galactic system}. Bull. Astron. Inst.
Netherl. \textbf{12}, No 458, 211-222.

Layzer, D. 1964. \textit{The formation of stars and galaxies: unified hypotheses}. Ann. Rev. Astron. Astrophys. \textbf{2}, 341-362.

Lebedinski, A.I. 1954. \textit{A hypothesis on star formation}. Problems of cosmogony \textbf{2}, 5-149 [\textit{in Russian}].

Ledoux, P. 1951. \textit{Sur la stabilite gravitationnelle d'une nebuleuse isotherme}. Ann. d'Astrophys. \textbf{14}, 438-447.

Lin, C.C. 1955. \textit{The theory of hydrodynamic stability}. Cambridge Univ. Press.

Lin, C.C. 1959. \textit{Hydrodynamics of liquid helium II}. Phys. Rev. Lett. \textbf{2}, 245-246.

Lin, C.C. 1966. \textit{On the mathematical theory of a galaxy of star.} J. SIAM Appl. Math. \textbf{14}, No 4, 876-920 (June 8-9,
1965 Courant Symp., Courant Inst. Math. Sci., NY).

Lin, C.C. 1967a. \textit{Stellar dynamical theory of normal spirals}. Lect. Appl. Math. \textbf{9}, 66-97 (1965 Summer school,
Cornell Univ. Ithaca, NY).

Lin, C.C. 1967b. \textit{The dynamics of disk-shaped galaxies}. Ann. Rev. Astron. Astrophys. \textbf{5}, 453-464.

Lin, C.C., Shu, F.H. 1964. \textit{On the spiral structure of disk galaxies}. ApJ \textbf{140}, 646-655 (LS64).

Lin, C.C., Shu, F.H. 1966. \textit{On the spiral structure of disk galaxies. II. Outline of a theory of density waves}. Proc. Nat. Acad. Sci. \textbf{55}, 229-234.

Lin, C.C., Shu, F.H. 1967. \textit{Density waves in disk galaxies}. In: Radio astronomy and the Galactic system
(Proc. IAU Symp. No 31, Noordwijk 1966), H. van Woerden, ed. London {\&} NY,
Academic Press, 313-317.

Lin, C.C., Yuan, C., Shu, F.H. 1969. \textit{On the structure of disk galaxies. III. Comparison with observations}. ApJ \textbf{155}, 721-746.

Lindblad, B. 1926a. \textit{Cosmogonic consequences of a theory of the stellar system}. Ark. \textbf{19A}, No 35.

Lindblad, B$. $1926b. \textit{Star-streaming and the structure of stellar systems. Paper 2}. Ark. \textbf{19B}, No 7.

Lindblad, B. 1927a. \textit{On the nature of the spiral nebulae}. MN \textbf{87}, 420-426.

Lindblad, B. 1927b.\textit{ On the state of motion in the galactic system}. MN \textbf{87}, 553-564.

Lindblad, B. 1927c. \textit{The small oscillations of a rotating stellar system and the development of spiral arms}. Ark. \textbf{20A}, No 10.

Lindblad, B. 1929. \textit{On the relation between the velocity ellipsoid and the rotation of the Galaxy}. Ark. \textbf{21A}, No 15.

Lindblad, B. 1934. \textit{The orientation of the planes of spiral nebulae inferred from the dark lanes of occulting matter}. Ark. \textbf{24A}, No 21.

Lindblad, B. 1938. \textit{On the theory of spiral structure in the nebulae}. Zeitschr. f. Astrophys. \textbf{15}, 124-136.

Lindblad, B. 1948. \textit{On the dynamics of stellar systems: (G. Darwin lecture)}. MN \textbf{108}, 214-235.

Lindblad, B. 1956. \textit{Contributions to the theory of spiral structure}. SOA \textbf{19}, No 7.

Lindblad, B. 1959. \textit{Galactic dynamics}. In: Handbuch der Physik, \textbf{53,} 21. S. Flugge, ed.
Berlin, Springer-Verlag.

Lindblad, B. 1961. \textit{On the formation of dispersion rings in the central layer of a galaxy}. SOA \textbf{21}, No 8.

Lindblad, B. 1962a. \textit{Interstellar Matter in Galaxies as Revealed by Observations in the Optical Spectral Regions}. In: Problems of Extragalactic Research (Proc. IAU
Symp. No15). G.C. McVittie, ed. NY, Macmillan Press, 57-69.

Lindblad, B. 1962b. \textit{Theories of spiral structure in galaxies}. In: Problems of Extragalactic Research (Proc. IAU
Symp. No 15). G.C. McVittie, ed. NY, Macmillan Press, 146-165.

Lindblad, B. 1963. \textit{On the possibility of a quasi-stationary spiral structure in galaxies}. SOA \textbf{22}, No 5.

Lindblad, B. 1964. \textit{On the circulation theory of spiral structure}. Astrophysica Norvegica \textbf{9}, 103-111.

Lindblad, B., Langebartel, R. 1953. \textit{On the dynamics of stellar systems}. SOA \textbf{17}, No 6.

Lindblad, B., Brahde R. 1946. \textit{On the direction of rotation in spiral nebulae}. ApJ \textbf{104}, 211-225.

Lindblad, P.O. 1962. \textit{Gravitational resonance effects in the central layer of a galaxy}. In: The distribution and motion of interstellar matter
in galaxies (Proc. Conf. Inst. Adv. Study, Princeton, 1961). L.Woltjer, ed.
W.A.Benjamin, NY, 222-233.

Lindblad, P.O., Lindblad, B. 1958. \textit{On rotating ring orbits in galaxies}. In: Comparison of the Large-Scale
Structure of the Galactic System with that of Other Stellar Systems (Proc.
IAU Symp. No 5, Dublin 1955). N.G. Roman, ed. Cambridge Univ. Press, 8-15.

Lynden-Bell, D. 1960a. \textit{Can spherical clusters rotate?} MN \textbf{120}, 204-213.

Lynden-Bell, D. 1960b. \textit{Stellar and galactic dynamics}. PhD thesis, Univ. of Cambridge.

Lynden-Bell, D. 1962. \textit{The stability and vibrations of a gas of stars}. MN \textbf{124}, 279-296.

Maxwell, J.C. 1859. \textit{On the stability of the motion of Saturn's rings.} Cambridge. (In: The scientific papers. Cambridge: Univ.
Press. 1890, \textbf{1}, 288).

Mayall, N.M., Aller, L.H. 1942. \textit{The rotation of the spiral nebula Messier 33}. ApJ \textbf{95}, 5-23.

Morgan, W.W., Sharpless, S., Osterbrock, D.E. 1952. \textit{Some features of Galactic structure in the neighborhood of the Sun.} Astron. J. \textbf{57},
3.

Morgan, W.W., Mayall, N.Y. 1957. \textit{A spectral classification of galaxies}. Publ. Astron. Soc. Pacific \textbf{69},
291-303.

Munch, G. 1959.\textit{ The mass-luminosity ratio in stellar systems}. Publ. Astron. Soc. Pacific \textbf{71}, 101- 105.

Ogorodnikov, K.F. 1958. \textit{Principles of stellar dynamics}. Moscow, GIFML Publ. [\textit{in Russian}].

Oort, J.H. 1962. \textit{Spiral structure}. In: The distribution and motion of interstellar matter in
galaxies (Proc. conf. Inst. Adv. Study, Princeton, 1961). L.Woltjer, ed.
W.A.Benjamin, NY, 234-244.

Oort, J.H. 1967.\textit{ Bertil Lindblad: (Obituary notice)}. Quart. J. Roy. Astron. Soc. \textbf{7}, 329-341.

Oort, J.H., Westerhout, G., Kerr F.  1958. \textit{The Galactic system as a spiral nebula.} MN \textbf{118}, 379-389.

Parenago, P.P. 1950. \textit{On the gravitational potential of the Galaxy}. юstron. Zh. \textbf{27}, 329.

Parenago, P.P. 1952. \textit{On the gravitational potential of the Galaxy}. юstron. Zh. \textbf{29}, 266.

Pasha, I.I. 2000. \textit{The fate of the galactic theories of James Jeans and Bertil Lindblad}. Istoriko-Astron. Issled. \textbf{25}, 91-116.

Prendergast, K.H. 1962. In: The distribution and motion of interstellar
matter in galaxies (Proc. Conf. Inst. Adv. Study, Princeton, 1961).
L.Woltjer, ed. W.A.Benjamin, NY, 318.

Prendergast, K.H., Burbidge, G.R. 1960. \textit{The persistence of spiral structure}. ApJ \textbf{131}, 243-246.

Reid, W.H., Lin, C.C. 1963. \textit{Turbulent flow, theoretical aspects}. Handbuch der Physik \textbf{VIII/2}.

Safronov, V.S. 1952. \textit{The material density in the solar neighborhood of the Galaxy}. юstron. Zh. \textbf{29}, 198.

Safronov, V.S. 1960. \textit{On the gravitational instability in flat axisymmetric rotating systems}. Dokl. Akad. Nauk SSSR \textbf{130}, 53-56.

Safronov, V.S. 1960. \textit{On the gravitational instability in flattened systems with axial symmetry and non-uniform rotation}. Ann. d'Astrophys. \textbf{23}, 979-982.

Sandage, A. 1961. \textit{The Hubble Atlas of Galaxies}. Carnegie Inst. of Washington.

Schatzman, E. 1954. \textit{Critical review of cosmogonic theories prevailing in West Europe and in America. Part 1}. Problems of cosmogony \textbf{3}, 227-308 [\textit{in Russian}].

Schmidt, M. 1956. \textit{A model of the distribution of mass in the Galactic system}. Bull. Astron. Netherl. \textbf{13}, No 468, 15-41.

Shu, F.H. 1964. \textit{Gravitational instability and spiral structure in disk galaxies}. B.Sc. Thesis in Physics, MIT.

Shu, F.H. 2001. Private communication.

Toomre, A. 1963. \textit{On the distribution of matter within highly flattened galaxies}. ApJ \textbf{138}, 385-392.

Toomre, A. 1964. \textit{On the gravitational stability of a disk of stars}. ApJ \textbf{139}, 1217-1238 (T64).

Toomre, A. 1977. \textit{Theories of spiral structure}. Ann. Rev. Astron. Astrophys. \textbf{15}, 437-478.

Toomre, A. 1996. \textit{Some historical remarks on Bertil Lindblad's work on galactic dynamics}. In: Barred galaxies and circumnuclear activity (Proc.
Nobel Symp. No 98), A. Sandqvist, P.O.Lindblad (Eds.)., 1-5.

Vaucouleurs, G. de 1958. \textit{Tilt criteria and direction of rotation of spiral galaxies}. ApJ \textbf{127}, 487-503.

Vaucouleurs, G. de 1959. \textit{Classification and morphology of external galaxies}. In: Handbuch der Physik \textbf{53}. S. Flugge,
ed. Berlin, Springer - Verlag, 275-310.

Vlasov, A.A. 1959. \textit{On the spatially inhomogeneous distributions in a system of gravitating particles} (Proc. 6$^{th}$ Conf. on the Problems of Cosmogony,
1957), Moscow, USSR Acad. Sci. Publ., 116-130 [\textit{in Russian}].

Vorontsov-Velyaminov, B.A. 1959. \textit{Morphological Catalog of Galaxies}, \textbf{1}. Moscow, Moscow State Univ.
Publ.

Vorontsov-Velyaminov, B.A. 1972. \textit{Extragalactic astronomy}. Moscow, Nauka Publ. [\textit{in Russian}].

Weizsacker, C. F. von 1951. T\textit{he evolution of galaxies and stars}. ApJ \textbf{114}, 165-186.

Zonn, W., Rudnicki,$ K. $1957. \textit{Astronomia gwiazdowa.} Panstwowe wydawnoctwo naukowe, Warszawa.

Zwicky, F. 1957. \textit{Morphological astronomy}. Berlin - Gottingen - Heidelberg. Springer - Verlag.
\par }
}

\end{document}